\documentclass[12pt, draftclsnofoot, onecolumn]{IEEEtran}
\usepackage{soul}            
\usepackage{multicol}
\usepackage{epsfig}
\usepackage{epstopdf}
\usepackage{float}
\usepackage{perpage}
\MakeSorted{table}
\usepackage{array}
\usepackage{relsize}
\usepackage{tikz}
\usepackage{enumitem}
\usepackage{graphicx}
\usepackage{amsfonts}
\usepackage{amssymb}

\usepackage{etoolbox}
\usepackage[noend]{algpseudocode}
\usepackage{cite}
\usepackage{tikz}
\DeclareMathAlphabet\mathbfcal{OMS}{cmsy}{b}{n}
\usepackage[cmex10]{amsmath}
\usepackage{array}
\usepackage{fixltx2e}
\usepackage{wrapfig}
\usepackage[hidelinks]{hyperref}
\usepackage{mathtools, cuted}
\usepackage{hhline}

\usepackage{tabularx}
\usepackage{subcaption,graphicx}
    \newcolumntype{L}{>{\raggedright\arraybackslash}X}
    \usepackage{pifont}
    
\ifCLASSINFOpdf
\else
\fi
\usepackage[font=footnotesize,labelfont=bf, figurename=Fig.]{caption}
\usepackage{subcaption}

\usepackage{etoolbox} 
\newtoggle{SINGLE_COL}
\toggletrue{SINGLE_COL}     

\makeatletter
\def\BState{\State\hskip-\ALG@thistlm}
\makeatother
\newcommand\norm[1]{\left\lVert#1\right\rVert}
\newcommand{\bnewsymbol}{
\begin{tikzpicture}
\filldraw[fill=black,draw=black] circle (4pt);
\end{tikzpicture}
}

\begin{document}
\bstctlcite{IEEEexample:BSTcontrol}
\title{Uplink Sum-Rate and Power Scaling Laws for Multi-User Massive MIMO-FBMC Systems}

\author{\IEEEauthorblockN{\iftoggle{SINGLE_COL}{Prem Singh, Himanshu B. Mishra, Aditya K. Jagannatham, K. Vasudevan, and Lajos Hanzo, \emph{Fellow, IEEE}}{Prem Singh, Himanshu B. Mishra, Aditya K. Jagannatham, K. Vasudevan, and Lajos Hanzo, \emph{Fellow, IEEE}}}
\thanks{L. Hanzo would like to acknowledge the financial support of the EPSRC/UK and the ERC's Advanced Fellow grant QualitCom.}

}

\IEEEtitleabstractindextext{%
\begin{abstract}
This paper analyses the performance of filter bank multicarrier (FBMC) signaling in conjunction with offset quadrature amplitude modulation (OQAM) in multi-user (MU) massive multiple-input multiple-output (MIMO) systems. Initially, closed form expressions are derived for tight lower bounds corresponding to the achievable uplink sum-rates for FBMC-based single-cell MU massive MIMO systems relying on maximum ratio combining (MRC), zero forcing (ZF) and minimum mean square error (MMSE) receiver processing with/without perfect channel state information (CSI) at the base station (BS). This is achieved by exploiting the statistical properties of the intrinsic interference that is characteristic of FBMC systems.  Analytical results are also developed for power scaling in the uplink of MU massive MIMO-FBMC systems. The above analysis of the achievable sum-rates and corresponding power scaling laws is subsequently extended to multi-cell scenarios considering both perfect as well as imperfect CSI, and the effect of pilot contamination. The delay-spread-induced performance erosion imposed on the linear processing aided BS receiver is numerically quantified by simulations. Numerical results are presented to demonstrate the close match between our analysis and simulations, and to illustrate and compare the performance of FBMC and traditional orthogonal frequency division multiplexing (OFDM)-based MU massive MIMO systems.
\end{abstract}

}

\maketitle

\IEEEdisplaynontitleabstractindextext

\IEEEpeerreviewmaketitle

\section{Introduction}
\IEEEPARstart{I}{n} recent years, massive multiple-input multiple-output (MIMO) technology \cite{marzetta2016fundamentals} has gained significant popularity due to its higher throughput and ability to simultaneously support a large number of users. Employing a large number of antennas (few hundred) enables the base station (BS) in such systems to suppress the co-channel interference using low-complexity linear receivers such as maximum ratio combining (MRC), zero forcing (ZF) and minimum mean square error (MMSE), which leads to a significant spectral efficiency improvement. Orthogonal frequency division multiplexing (OFDM), which circumvents the degradation resulting from the frequency selective nature of wireless channels, has recently been applied in massive MIMO systems \cite{PitarokoilisBL16,Vasu2017Massive}. However, the rectangular time-domain pulse of OFDM leads to a sinc-shaped out-of-band (OOB) emission. Furthermore, the ability of OFDM to partition the wideband spectrum into multiple sub-bands of orthogonal subcarriers requires accurate frequency- and timing-synchronization of the multiple users within the cyclic prefix (CP) duration. OFDM systems are thus sensitive to synchronization errors such as carrier frequency offset (CFO) \cite{PolletBM95}, especially in the uplink, where it is challenging to track the Doppler shifts of different users~\cite{MorelliKP07}.

The OFDM systems relying on offset quadrature amplitude modulation (OQAM) (popularly known as OQAM based filter bank multicarrier (FBMC) systems) \cite{Farhang-Boroujeny16}, which allow the introduction of an efficient sharp pulse shaping filter, exhibit a lower OOB radiation than classic CP-OFDM. These beneficial pulse shaping filters alleviate the stringent uplink synchronization requirements of FBMC-OQAM systems and eliminate the need for CP that is required to combat inter-symbol-interference (ISI) in classic OFDM systems \cite{AminjavaheriFRF15,siohan2002analysis}. This leads to an improved spectral efficiency in FBMC-OQAM systems. The advantages of FBMC over OFDM in the context of cognitive radios and the uplink of multi-user (MU) networks have recently been studied in \cite{Farhang-Boroujeny08} and \cite{SourckWBSF11}, respectively. In light of the aforementioned advantages, FBMC-OQAM systems are being  considered as potential waveform candidates to replace OFDM in next-generation wireless cellular systems \cite{NisselSR17,farhang2011ofdm,cherubini2000filter}.  Recently, the use of FBMC-OQAM transmission has been extended to both MIMO  \cite{Perez-NeiraCZRK16} and massive MIMO systems \cite{RottenbergMHL18}. The focus of this paper is therefore to design and analyse the performance of MU massive MIMO systems based on FBMC-OQAM signaling. For brevity, FBMC-OQAM is simply referred to as FBMC in the sequel.
\subsection{Review of Existing Works}
In contrast to OFDM, the OQAM based FBMC adopts real OQAM symbols since the orthogonality holds in the real field only \cite{siohan2002analysis}. The resulting \emph{intrinsic interference} renders amalgamation of FBMC with massive MIMO systems challenging \cite{SinghRMV18}. Hence, it is not always possible to extend the existing analysis of OFDM-based massive MIMO systems to that of the massive MIMO-FBMC systems. Thus, the performance analysis of FBMC-based massive MIMO techniques warrants meticulous investigation. There are some studies in the existing literature that have investigated the application of FBMC in the context of massive MIMO systems. For instance, the authors of \cite{AminjavaheriFDF17} demonstrate that the signal to noise-plus-interference ratio (SINR) of frequency selective single-cell massive MIMO-FBMC systems is limited by a deterministic value governed by the correlation between the multi-antenna combine tap weights and the channel impulse responses. An equalizer is designed in \cite{AminjavaheriFF18} that removes the correlation induced SINR-limitation described in \cite{AminjavaheriFDF17}. References \cite{MestreRN16,RottenbergMHL18} theoretically characterize the mean squared error (MSE) of the estimated symbols in the uplink of a single-cell massive MIMO-FBMC system relying on linear receivers such as ZF, MMSE and matched filtering. The authors of \cite{FarhangMDF14,AminjavaheriFMD15} have compared FBMC and CP-OFDM schemes in the context of single-cell massive MIMO systems, indicating several benefits over the latter such as reduced complexity, lower sensitivity to CFO, reduction of peak-to-average power ratio, reduced latency and increased bandwidth efficiency. 
The above studies reflect that FBMC has indeed attracted significant research interests and it is a compelling signalling technique in combination with massive MIMO for next generation wireless systems. All the works reviewed above are restricted to single-cell massive MIMO-FBMC systems. Furthermore, they rely on the idealized simplifying assumption of having perfect channel state information (CSI) at the BS. To the best of our knowledge, the achievable uplink sum-rates of single- and multi-cell massive MIMO systems using FBMC signaling for transmission over quasi-static channels in the presence of both perfect and imperfect CSI at the BS have not been disseminated in the open literature. This paper aims to fill this void in the existing literature on FBMC-based MU massive MIMO systems.

\subsection{Contributions of Present Work}
The analysis of the uplink of FBMC-based MU massive MIMO systems is quite challenging due to the following constraints imposed on FBMC signaling in contrast to its OFDM counterpart.
i) The virtual FBMC symbols obtained at the output of the FBMC receive filter bank comprise both the original OQAM symbol and the resultant intrinsic interference. Thus, the statistical properties of the intrinsic interference have to be shown for deriving the analytical results for the uplink of FBMC-based massive MIMO systems. ii)  The preprocessing step invoked for facilitating the OQAM to QAM conversion at the BS poses significant challenges in terms of determining the statistical characteristics of the noise pulse interference at the output of linear receivers. Additionally, the noise plus interference arising during the OQAM to QAM conversion also has to be analysed for obtaining the eventual SINR expression for the various receivers, both in single- as well multi-cell scenarios. iii) The channel estimation in massive MIMO-FBMC systems requires the insertion of zero symbols between the adjacent training symbols to avoid ISI that arises due to the overlapping nature of the time domain FBMC symbols. This, in turn, requires separate analysis for the resultant intrinsic interference to compute the virtual training symbols for purpose of the channel estimation. Furthermore, the OQAM training symbols have to be precoded at the transmitter for ensuring that the virtual training matrix at the receiver becomes orthogonal \cite{kofidis2013preamble}. Given the above challenges, our key contributions can be briefly summarized as~follows:
\begin{itemize} 
\item The analysis begins by determining the second-order statistical properties of the intrinsic interference, followed by the achievable ergodic uplink sum-rates for single-cell MU massive MIMO-FBMC systems relying on MRC, ZF and MMSE processing at the BS in the presence of both perfect as well as imperfect CSI.
\item Closed form expressions are derived for the lower bounds on the achievable uplink sum-rates for single-cell MU massive MIMO-FBMC systems relying on linear receiver processing at the BS both with perfect and imperfect CSI, followed by the corresponding power scaling laws.
\item The above sum-rate analysis is then extended to FBMC-based multi-cell MU massive MIMO systems, incorporating also the effect of imperfect CSI. The pertinent power scaling laws of this scenario are also determined.
\item The real field orthogonality of FBMC systems progressively degrades upon increasing the channel's dispersion. To study this effect, the impact of the channel's delay spread on the uplink performance of FBMC-based single- and multi-cell massive MIMO systems is quantified numerically. Furthermore, the effect of carrier frequency offset (CFO) on the uplink of FBMC and OFDM-based single- and multi-cell massive MIMO systems is also quantified numerically.
\item Simulation results validate the analytical expressions and also compare the performance of FBMC and OFDM-based massive MIMO systems.
\end{itemize}
\subsection{Organization and Notation of Paper}
The remainder of this paper is organized as follows. The next section presents the equivalent baseband model of our MU massive MIMO-FBMC system operating in a multipath fading channel. Section-\ref{SC_MU_FBMC} presents our analytical results for the FBMC-based single-cell MU massive MIMO systems both in the presence of perfect and imperfect receive CSI. Section-\ref{MC_MU_MMIMO} extends the analysis to FBMC-based multi-cell MU massive MIMO systems with/ without perfect CSI at the BS. Our simulation results are provided in Section-\ref{Results} and Section-\ref{Conclusion} concludes the paper.

\emph{Notation:} Upper and lower case bold face letters $\mathbf{A}$ and $\mathbf{a}$ denote matrices and vectors respectively. The superscripts $(\cdot)^{\ast}$, $(\cdot)^{T}$ and $(\cdot)^{H}$ represent the complex conjugate, transpose and Hermitian operators, respectively. The operators $\mathbb{\mathbb{E}}[\cdot]$ and $\text{Var}[\cdot]$ denote the expectation and variance, respectively, while $\text{Tr}(\cdot)$ and $\ast$ represent trace and convolution operators, respectively.  Further, $j \triangleq \sqrt{-1}$, $\Re \{\cdot\}$ and $\Im \{\cdot\}$ represent real and imaginary parts, and $\mathbf{I}_{M}$ represents the $M\times M$ identity matrix. Furthermore, diag$(\bar{a})$ represents a diagonal matrix with $\bar{a}$ on its principal diagonal and the notation $X\sim \mathcal{CN}(0,\sigma^{2})$ describes a zero-mean circularly symmetric complex Gaussian random variable $X$ with mean zero and variance $\sigma^{2}$.

\section{MU massive MIMO-FBMC System}\label{System_Model}
We consider the uplink of an FBMC-based MU massive MIMO system having $M$ subcarriers, with $U$ single-antenna users transmitting their signals in same time-frequency resources to a BS equipped with an array of $N$ antennas. Let  $d^{u}_{m,k}$ denote a real OQAM symbol of the $u$th user at subcarrier index $m$ and symbol instant $k$, which is generated by extracting the real and imaginary parts of the complex QAM symbol $c^{u}_{m,k}$ according to the rules described in \cite[Eq. (2), (3)]{viholainen2009prototype}.
Let $T$ represent the duration of the QAM symbol $c^{u}_{m,k}$ with $\frac{T}{2}$ denoting the duration of an OQAM symbol $d^{u}_{m,k}$.  The real and imaginary parts of the QAM symbol  $c^{u}_{m,k}$  are assumed to be spatially and temporally independent and identically distributed (i.i.d) with power $P_{d}$ such that $\mathbb{E}\big[d^{u}_{m,k}\left(d^{u}_{m,k}\right)^{\ast}\big]=P_{d}$. Hence, it follows that $\mathbb{E}\big[c^{u}_{m,k}\left(c^{u}_{m,k}\right)^{\ast}\big]=2P_{d}$. The equivalent discrete-time baseband FBMC transmit signal $s^{u}[l]$ of the $u$th user is expressed as \cite{siohan2002analysis}
 \begin{eqnarray}\label{eq:s_u}
s^{u}[l]=\sum_{m=0}^{M-1} \sum_{k\in\mathbb{Z}}^{}d^{u}_{m,k}\chi_{m,k}[l],\ \ \text{for}\ 1\leq u\leq U,
\end{eqnarray}
where $l$ denotes the sample index corresponding to the sampling rate $M/T$ and the basis function 
\begin{eqnarray}\label{eq:basis}
 \chi_{m,k}[l]&=& p\big[l-k{M}/2\big]e^{j{2\pi}{ml}/M}e^{j\phi_{m,k}}.
\end{eqnarray}
The phase factor $\phi_{m,k}$ above is defined as $\phi_{m,k}=\frac{\pi}{2}(m+k)-\pi mk$ \cite{siohan2002analysis}. The symmetric real-valued pulse $p[l]$ of length $L_{p}$ represents the impulse response of the \textit{prototype filter} of the FBMC system. The key differences between OFDM and FBMC systems lie i) in the fact that the latter adopts OQAM symbols rather than QAM symbols; and ii) in the specific choice of the prototype filter $p[l]$. The OFDM symbols are shaped using a time-domain  rectangular window that has a sinc-shaped spectrum resulting in OOB emissions. In order to overcome this impediment, the prototype pulse $p[l]$ in FBMC systems is well FT localised such that the basis function $\chi_{m,k}[l]$ satisfies the real field orthogonality condition $\Re\big\{\sum_{l=-\infty}^{+\infty}\chi_{m,k}[l]\chi_{\bar{m},\bar{k}}^{*}[l]\big\}=\delta_{m,\bar{m}}\delta_{k,\bar{k}}$ \cite{siohan2002analysis},
where $\delta_{m,\bar{m}}$ denotes the Kronecker delta with $\delta_{m,\bar{m}}=1$ if $m=\bar{m}$ and zero otherwise.
Let the quantity $\xi^{\bar{m},\bar{k}}_{m,k}$ be defined as $\xi^{\bar{m},\bar{k}}_{m,k}=\sum_{l=-\infty}^{+\infty}\chi_{m,k}[l]\chi_{\bar{m},\bar{k}}^{*}[l]$. Thus, we have $\xi^{\bar{m},\bar{k}}_{m,k}=1$ if $(m,k)=(\bar{m},\bar{k})$, and $\xi^{\bar{m},\bar{k}}_{m,k}=j\langle\xi\rangle^{\bar{m},\bar{k}}_{m,k}$ if $(m,k)\neq(\bar{m},\bar{k})$, where the quantity $\langle\xi\rangle^{\bar{m},\bar{k}}_{m,k}=\Im\{\sum_{l=-\infty}^{+\infty}\chi_{m,k}[l]\chi_{\bar{m},\bar{k}}^{*}[l]\}$ denotes the imaginary part of the cross-correlation between two basis functions \cite{lele2008channel}.

Let $g^{n,u}[l]$, for $0\leq l\leq L-1$, denotes an $L$-tap dispersive multipath fading channel between the $u$th user and the $n$th BS antenna. The signal received at the $n$th BS antenna can be obtained~as
\begin{eqnarray}\label{eq:Rec_Sig}
 y^{n}[l]=\sum_{u=1}^{U}\Big(s^{u}[l]\ast g^{n,u}[l]\Big)+ \eta^{n}[l],\ \ \text{for} \ \ 1\leq n\leq N,
\end{eqnarray}
where $\eta^{n}[l]$ represents the zero mean additive white Gaussian noise with power $\sigma^{2}_{\eta}$. The demodulated signal $y^{n}_{\bar{m},\bar{k}}$ on the $n$th BS antenna at subcarrier $\bar{m}$ and symbol time $\bar{k}$ is obtained via matched filtering with the FBMC basis function $\chi_{\bar{m},\bar{k}}[l]$ as $y^{n}_{\bar{m},\bar{k}}=\sum_{l=-\infty}^{+\infty}  y^{n}[l] \chi_{\bar{m},\bar{k}}^{\ast}[l]$. By substituting the expressions for $\chi_{\bar{m},\bar{k}}[l]$ and $y^{n}[l]$ from (\ref{eq:basis}) and (\ref{eq:Rec_Sig}) respectively, and assuming that the channel is quasi-static in nature with frequency flat fading across each subcarrier, i.e. that $P[l-i-kM/2]\approx P[l-kM/2]$ for $i\in[0,L]$ \cite{lele2008channel,choi2017pilot,kofidis2013preamble}- which is characteristic of FBMC systems- the expression for the demodulated signal $y^{n}_{\bar{m},\bar{k}}$ can be written similar to \cite{8561228,8727898} as
\begin{eqnarray}\label{eq:MIMO_FD1}
 y^{n}_{\bar{m},\bar{k}}= \sum_{u=1}^{U} G^{n,u}_{\bar{m}}\ b^{u}_{\bar{m},\bar{k}}+\eta^{n}_{\bar{m},\bar{k}},
\end{eqnarray}
where $G^{n,u}_{\bar{m}}$ denotes the CFR of the linear spanning from the $u$th user to the $n$th BS  antenna at the $\bar{m}$th subcarrier, and is determined as $G^{n,u}_{\bar{m}}=\sum_{l=0}^{L-1}g^{n,u}[l]e^{-j2\pi \bar{m}l/M}$. The demodulated noise $\eta^{n}_{\bar{m},\bar{k}}$ at the $n$th BS antenna is expressed as $ \eta^{n}_{\bar{m},\bar{k}}= \sum_{l=-\infty}^{+\infty}  \eta^{n}[l] \chi_{\bar{m},\bar{k}}^{\ast}[l]$, and is also distributed as $\mathcal{CN}(0,\sigma^{2}_{\eta})$ due to the linear demodulation operation. The quantity $b^{u}_{\bar{m},\bar{k}} = d^{u}_{\bar{m},\bar{k}}+j I^{u}_{\bar{m},\bar{k}}$ given by the addition of the OQAM symbol $ d^{u}_{\bar{m},\bar{k}}$ and the imaginary intrinsic interference component $I^{u}_{\bar{m},\bar{k}}$ can be considered to be the virtual symbol at the FT index $(\bar{m},\bar{k})$. Thus, it is necessary to determine the statistical properties of the intrinsic interference term $I^{u}_{\bar{m},\bar{k}}$ in order to obtain the SINR, the achievable rate and the lower bound expressions for the FBMC-based massive MIMO system. The interference $I^{u}_{\bar{m},\bar{k}}$ is expressed as 
\begin{eqnarray}\label{eq:Intrf}
I^{u}_{\bar{m},\bar{k}} = \sum_{\substack{(m,k)\in \Omega_{\bar{m},\bar{k}}}} d^{u}_{m,k}\langle\xi\rangle^{\bar{m},\bar{k}}_{m,k},
\end{eqnarray}
where $\Omega_{\bar{m},\bar{k}}$ denotes the neighbourhood of the desired FT point $(\bar{m},\bar{k})$ that does not include the point $(\bar{m},\bar{k})$\footnote{For well FT localized filters such as isotropic orthogonal transform algorithm (IOTA), a significant portion of the interference can be {attributed} to the first order neighbourhood of $(\bar{m},\bar{k})$, denoted by $\Omega_{\bar{m},\bar{k}}$ = $\left\{(\bar{m}\pm 1,\bar{k}\pm 1),(\bar{m},\bar{k}\pm 1),(\bar{m}\pm 1,\bar{k})\right\}$.}. The term $I^{u}_{\bar{m},\bar{k}}$ comprises both the ISI and the inter-carrier-interference (ICI) imposed by the symbols in the neighbourhood of the desired symbol at the index $(\bar{m},\bar{k})$. This is different from OFDM systems wherein the ISI is suppressed by using the CP, while the ICI is nulled due to the orthogonality of the subcarriers \cite{vasudevanIARIA}. The term $I^{u}_{\bar{m},\bar{k}}$ has a mean of zero and variance of
\begin{align}\label{eq:Var_Intr}
\mathbb{\mathbb{E}}[|I^{u}_{\bar{m},\bar{k}}|^{2}]\approx P_{d}.
\end{align}
A detailed proof of the above result is given in Appendix-\ref{Intrf_Cal}. Exploiting the above result and the property that the desired symbol $d^{u}_{\bar{m},\bar{k}}$ and the interference $I^{u}_{\bar{m},\bar{k}}$ are zero-mean independent variables, the variance of the virtual symbol $b^{u}_{\bar{m},\bar{k}}=d^{u}_{\bar{m},\bar{k}}+j{I}^{u}_{\bar{m},\bar{k}}$ can now be computed as $\mathbb{\mathbb{E}}[|b^{u}_{\bar{m},\bar{k}}|^{2}] =\mathbb{E}[|d^{u}_{\bar{m},\bar{k}}|^{2}]+ \mathbb{E}[|I^{u}_{\bar{m},\bar{k}}|^{2}]\approx 2P_{d}$. For convenience, (\ref{eq:MIMO_FD1}) can be succinctly represented in vector form as
\begin{eqnarray}\label{eq:MIMO_FD}
\mathbf{y}_{\bar{m},\bar{k}}= \mathbf{G}_{\bar{m}}\mathbf{b}_{\bar{m},\bar{k}}+ \boldsymbol{\eta}_{\bar{m},\bar{k}},
\end{eqnarray}
where $\mathbf{y}_{\bar{m},\bar{k}}=[y^{1}_{\bar{m},\bar{k}},y^{2}_{\bar{m},\bar{k}},\ldots,y^{N}_{\bar{m},\bar{k}}]^{T}\in\mathbb{C}^{N\times 1}$ is the concatenated vector of received symbols at the BS across the $N$ antennas and $\boldsymbol{\eta}_{\bar{m},\bar{k}}=[\eta^{1}_{\bar{m},\bar{k}},\eta^{2}_{\bar{m},\bar{k}}, \ldots ,\eta^{N}_{\bar{m},\bar{k}}]^{T}\in\mathbb{C}^{N\times 1}$ is the noise vector with the covariance matrix $\mathbb{E}[\boldsymbol{\eta}_{\bar{m},\bar{k}}\boldsymbol{\eta}_{\bar{m},\bar{k}}^{H}]=\sigma^{2}_{\eta}\mathbf{I}_{N}$.  The vector $\mathbf{b}_{\bar{m},\bar{k}}=[b^{1}_{\bar{m},\bar{k}}, b^{2}_{\bar{m},\bar{k}},\ldots,b^{U}_{\bar{m},\bar{k}}]^{T}\in\mathbb{C}^{U\times 1}$ comprises the virtual symbols for all the $U$ users with the covariance matrix $\mathbb{\mathbb{E}}[\mathbf{b}_{\bar{m},\bar{k}}\mathbf{b}^{H}_{\bar{m},\bar{k}}]\approx2P_{d}\mathbf{I}_{U}$. The matrix $\mathbf{G}_{\bar{m}}=[\mathbf{g}^{1}_{\bar{m}},\mathbf{g}^{2}_{\bar{m}},\ldots,\mathbf{g}^{U}_{\bar{m}}]\in\mathbb{C}^{N\times U}$ is the CFR matrix on the $\bar{m}$th subcarrier between the BS and the $U$ users in the MU massive MIMO setup. The matrix $\mathbf{G}_{\bar{m}}$ is typically modelled as \cite{NgoLM13}
\begin{eqnarray}\label{eq:G_Mtx}
\mathbf{G}_{\bar{m}}=\mathbf{H}_{\bar{m}} \big[\text{diag}(\beta^{1},\beta^{2},\ldots,\beta^{U})\big]^{1/2}= \mathbf{H}_{\bar{m}}\mathbf{D}^{1/2},
\end{eqnarray}
where $\beta^{u}$ denotes the large-scale fading coefficient for user $u$ and the diagonal matrix $\mathbf{D}=\text{diag}(\beta^{1},\beta^{2},\ldots,\beta^{U})\in\mathbb{R}^{U\times U}$. The quantity $\beta^{u}$, which is constant over many coherence time intervals, is assumed to be independent over the BS antenna index $n$ and the subcarrier index $\bar{m}$, and known a priori. The matrix $\mathbf{H}_{\bar{m}}=[\mathbf{h}^{1}_{\bar{m}},\mathbf{h}^{2}_{\bar{m}},\ldots,\mathbf{h}^{U}_{\bar{m}}]\in\mathbb{C}^{N\times U}$ comprises  the fading coefficients at the $\bar{m}$th subcarrier between the BS and the $U$ users. The elements of the matrix $\mathbf{H}_{\bar{m}}$ are modeled as i.i.d. $\mathcal{CN}(0,1)$. Thus, for simplicity of analysis, the channel matrix $\mathbf{G}_{\bar{m}}=\mathbf{H}_{\bar{m}}\mathbf{D}^{1/2}$ is assumed to be spatially uncorrelated similar to the contributions such as \cite{AminjavaheriFDF17,AminjavaheriFF18}. After receiver processing at the BS, the estimate $\hat{\mathbf{c}}_{\bar{m},\bar{k}}=[\hat{c}^{1}_{\bar{m},\bar{k}},\ldots,\hat{c}^{U}_{\bar{m},\bar{k}}]^{T}\in\mathbb{C}^{U\times1 }$ of the transmitted QAM symbol vector is reconstructed from the estimated OQAM symbol vector $\hat{\mathbf{d}}_{\bar{m},\bar{k}}=[\hat{d}^{1}_{\bar{m},\bar{k}},\ldots, \hat{d}^{U}_{\bar{m},\bar{k}}]^{T}\in\mathbb{C}^{U\times1 }$  as~\cite[Eq. (7)]{viholainen2009prototype}
\begin{eqnarray}\label{eq:R2C}
\hat{\mathbf{c}}_{\bar{m},\bar{k}}=\left\{
  \begin{array}{@{}ll@{}}
    \hat{\mathbf{d}}_{\bar{m},2\bar{k}}+j\hat{\mathbf{d}}_{\bar{m},2\bar{k}+1}, & \bar{m}\ \text{even} \\
   \hat{\mathbf{d}}_{\bar{m},2\bar{k}+1}+j\hat{\mathbf{d}}_{\bar{m},2\bar{k}}, &\bar{m}\ \text{odd}.
  \end{array}\right.
\end{eqnarray}

\section{Single-Cell MU Massive MIMO-FBMC System}\label{SC_MU_FBMC}
Let $\mathbf{A}_{\bar{m}}\in\mathbb{C}^{N\times U}$ denote the combiner matrix employed at the BS. The estimate of the $U\times 1$ OQAM symbol vector at the output of the combiner is obtained as $\mathbf{\hat{d}}_{\bar{m},\bar{k}}=\Re\big\{\mathbf{A}_{\bar{m}}^{H}\mathbf{y}_{\bar{m},\bar{k}}\big\}$. The combiner matrix $\mathbf{A}_{\bar{m}}$ for the MRC, ZF and MMSE receivers, which are frequently employed in literature due to their linear nature and low complexity, is expressed as
\begin{eqnarray}\label{eq:Combiners}
\mathbf{A}_{\bar{m}}=
 \left\{
  \begin{array}{@{}ll@{}}
    \mathbf{G}_{\bar{m}}& \text{for}\ \text{MRC} \\
   \mathbf{G}_{\bar{m}}\big(\mathbf{G}_{\bar{m}}^{H}\mathbf{G}_{\bar{m}}\big)^{-1}& \text{for}\ \text{ZF}\\
   \bigg(\mathbf{G}_{\bar{m}}\mathbf{G}_{\bar{m}}^{H}+\dfrac{\sigma^{2}_{\eta}}{2P_{d}}\mathbf{I}_{N}\bigg)^{-1}\mathbf{G}_{\bar{m}}& \text{for}\ \text{MMSE}.
  \end{array}\right.
\end{eqnarray}
In subsequent sections, we derive the ergodic uplink sum-rates and the corresponding lower bounds, and the power scaling laws for the aforementioned receivers considering the operating regime where $1\ll U\ll N$ \cite{NgoLM13}. The following results will be used in the ensuing analysis.

 Let $\mathbf{a}=[a_{1},\ldots,a_{N}]^{T}$ and $\mathbf{b}=[b_{1},\ldots,b_{N}]^{T}$ be the $N\times 1$ mutually independent random vectors, which consist of zero mean i.i.d.  elements with variance $\sigma^{2}_{a}$ and $\sigma^{2}_{b}$, respectively. Then, from law of large numbers, it can be shown that \cite{cramer2004random}
\begin{eqnarray}\label{eq:Prel1}
\dfrac{1}{N}\mathbf{a}^{H}\mathbf{a}\xrightarrow[N\rightarrow \infty]{\text{a.s.}}\sigma^{2}_{a}\ \text{and} \dfrac{1}{N}\mathbf{a}^{H}\mathbf{b}\xrightarrow[N\rightarrow \infty]{\text{a.s.}}0,
\end{eqnarray}
where $\xrightarrow[N\rightarrow \infty]{\text{a.s.}}$ denotes almost sure convergence as $N\rightarrow \infty$. Furthermore,
\begin{eqnarray}\label{eq:Prel2}
\dfrac{1}{\sqrt{N}}\mathbf{a}^{H}\mathbf{b}\xrightarrow[N\rightarrow \infty]{\text{d}}\mathcal{CN}(0,\sigma^{2}_{a}\sigma^{2}_{b}),
\end{eqnarray}
where $\xrightarrow[N\rightarrow \infty]{\text{d}}$ denotes convergence in distribution as $N\rightarrow \infty$. Finally, the result below holds for two complex random matrices $\mathbf{X}$ and $\mathbf{Y}$ \cite{larsson2008space}
\begin{eqnarray}\label{eq:Prel3}
\mathbb{E}\big[\Re\{\mathbf{X}\}\Re\{\mathbf{Y}\}\big]=\dfrac{1}{2}\Re\big\{\mathbb{E}[\mathbf{XY}+\mathbf{XY^{\ast}}]\big\}.
\end{eqnarray}
The next subsection presents the sum-rate analysis for a single-cell MU massive MIMO-FBMC systems with imperfect CSI at the BS. The corresponding results for the perfect CSI are subsequently derived as a special~case.
\subsection{Imperfect CSI}\label{SC_MU_FBMC_IMCSI}
In practice, the channel matrix $\mathbf{G}_{\bar{m}}$ in a MU massive MIMO-FBMC system is estimated at the BS using uplink training symbols as described below.
\begin{figure}[!t]
\centering
\includegraphics [scale=0.73]{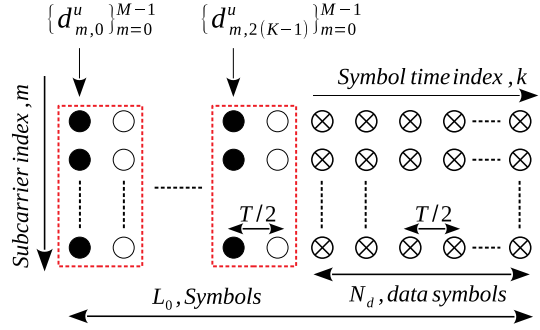}
\caption{Frame structure for the $u$th user. The symbols, \protect\bnewsymbol, $\bigcirc$ and $\bigotimes$ represent the training, zero and data symbols, respectively.}
\label{fig:frame}
\end{figure}
\subsubsection{Training-based linear MMSE Channel Estimation}
Consider $L_{0}$ OQAM symbols to be transmitted by the $u$th user on each subcarrier as per the frame structure illustrated in Fig.~\ref{fig:frame}. Let each frame comprises $K(K\geq U)$ training symbols to be employed for channel estimation, followed by $N_{d}$ data-bearing symbols. Since the adjacent FBMC symbols interfere with each other in the time domain due to the overlapping nature of the pulse-shaping filters, a zero symbol is inserted between the adjacent training symbols for reducing ISI to an acceptable level  \cite{lele2008channel,du2009novel,kofidis2013preamble,
 kong2013frequency}, as shown in Fig.~\ref{fig:frame}. In view of the inter-frame time gap commonly used in wireless communication, insertion of a zero symbol at the beginning of the frame is in general unnecessary \cite{kofidis2013preamble}. Thus, MIMO-FBMC pilot sequences with guard (zero) symbols require $2K$ OQAM symbols on each subcarrier, which is equivalent to $K$ complex QAM symbols \cite{kofidis2013preamble}. Hence, the training overhead required for channel estimation in MIMO-FBMC is similar to that of MIMO-OFDM \cite{BigueshG06} and does not incur any additional loss in spectral efficiency.

Evaluating (\ref{eq:MIMO_FD}) at the training symbol locations $k=2i$ for $0\leq i\leq K-1$ and stacking the resulting outputs, one obtains
\begin{eqnarray}\label{eq:CE_MODEL}
\mathbf{Y}_{\bar{m}}=\mathbf{G}_{\bar{m}}\mathbf{B}_{\bar{m}}^{T}+\mathbf{W}_{\bar{m}}=\sum_{j=1}^{U}\mathbf{g}^{j}_{\bar{m}}(\mathbf{b}^{j}_{\bar{m}})^{T}+\mathbf{W}_{\bar{m}},
\end{eqnarray}
where $\mathbf{Y}_{\bar{m}}=[\mathbf{y}_{\bar{m},0},  \mathbf{y}_{\bar{m},2},\ldots, \mathbf{y}_{\bar{m},2(K-1)}]\in\mathbb{C}^{N\times K}$ is the matrix of concatenated receive training vectors and $\mathbf{W}_{\bar{m}}=[\boldsymbol{\eta}_{\bar{m},0},  \boldsymbol{\eta}_{\bar{m},2},\ldots, \boldsymbol{\eta}_{\bar{m},2(K-1)}]\in\mathbb{C}^{N\times K}$ is the corresponding noise matrix. Each element of the noise matrix $\mathbf{W}_{\bar{m}}$ is distributed as $\mathcal{CN}(0,\sigma^{2}_{\eta})$.  The virtual training matrix $\mathbf{B}_{\bar{m}}=[\mathbf{b}^{1}_{\bar{m}},\mathbf{b}^{2}_{\bar{m}},\ldots, \mathbf{b}^{U}_{\bar{m}}]\in\mathbb{C}^{K\times U}$ is obtained by concatenation of the virtual training vectors, where the training vector for the $u$th user is $\mathbf{b}^{u}_{\bar{m}}=[b^{u}_{\bar{m},0},b^{u}_{\bar{m},2},\ldots, b^{u}_{\bar{m},2(K-1)}]^{T}\in\mathbb{C}^{K\times 1}$. The $i$th element of $\mathbf{b}^{u}_{\bar{m}}$ at the FT index $(\bar{m},2i)$ is given as
$b^{u}_{\bar{m},2i}=d^{u}_{\bar{m},2i}+jI^{u}_{\bar{m},2i}$.
The intrinsic interference $I^{u}_{\bar{m},2i}$, for $0\leq i\leq K-1$ and user $u$, can be expressed as
\begin{align}
\iftoggle{SINGLE_COL}{}{\nonumber &} I^{u}_{\bar{m},2i}=\sum_{\substack{ m\neq \bar{m}}} d^{u}_{m,2i}\Im\Big\{\sum_{l=-\infty}^{+\infty} p^{2}[l]e^{j(\phi_{m,0}-\phi_{\bar{m},0})} \iftoggle{SINGLE_COL}{}{\\ &}
e^{j{2\pi}(m-\bar{m})l/M}\Big\}=
\sum_{\substack{ m\neq \bar{m}}} d^{u}_{m,2i}\langle\xi\rangle^{\bar{m},0}_{m,0}.
\end{align}
A detailed proof of the above result is given in Appendix-\ref{Inter_Cal_chan}. Similar to \cite{singh2018cfo}, the training symbols are generated by extracting the real and imaginary parts of the random complex QAM symbols. Thus, for an orthogonal training matrix $\mathbf{B}_{\bar{m}}$ \cite{kofidis2013preamble}, constructed as per the procedure in Appendix-\ref{Othr_Bm}, it follows from \eqref{eq:Var_Intr} that $\mathbf{B}^{H}_{\bar{m}}\mathbf{B}_{\bar{m}}=P_{p}\mathbf{I}_{U}$, where $P_{p}=2P_{d}K$ represents the pilot power. The $N\times 1$ received training vector at the $\bar{m}$th subcarrier for the $u$th user can be obtained using \eqref{eq:CE_MODEL}~as
\begin{eqnarray}\label{eq:MMSE_CE_1}
\mathbf{y}^{u}_{\bar{m}}=\mathbf{Y}_{\bar{m}}(\mathbf{b}^{u}_{\bar{m}})^{\ast}=P_{p}\mathbf{g}^{u}_{\bar{m}}+\mathbf{w}_{\bar{m}}^{u}.
\end{eqnarray}
Here we have exploited the property that $(\mathbf{b}^{j}_{\bar{m}})^{T}(\mathbf{b}^{u}_{\bar{m}})^{\ast}=P_{p}$ for $j=u$ and zero otherwise. The noise vector obeys $\mathbf{w}_{\bar{m}}^{u}=\mathbf{W}_{\bar{m}}(\mathbf{b}_{\bar{m}}^{u})^{\ast}$. Utilizing \eqref{eq:Var_Intr}, $\mathbb{E}[\mathbf{w}_{\bar{m}}^{u}(\mathbf{w}_{\bar{m}}^{u})^{H}]=P_{p}\sigma^{2}_{\eta}\mathbf{I}_{N}$. From \eqref{eq:MMSE_CE_1}, the estimate of the channel vector at the $\bar{m}$th subcarrier between the BS and the $u$th user is 
\begin{eqnarray}\label{eq:MMSE_CE_2}
\nonumber \mathbf{\hat{g}}^{u}_{\bar{m}}=\dfrac{\beta^{u}}{P_{p}\beta^{u}+\sigma^{2}_{\eta}}\mathbf{y}^{u}_{\bar{m}}.
\end{eqnarray}
It can be verified that the covariance matrix of $\mathbf{\hat{g}}^{u}_{\bar{m}}$ and the error vector $\mathbf{e}^{u}_{\bar{m}}=\mathbf{g}^{u}_{\bar{m}}-\mathbf{\hat{g}}^{u}_{\bar{m}}$~are
\begin{eqnarray}
\mathbb{E}[\mathbf{\hat{g}}^{u}_{\bar{m}}(\mathbf{\hat{g}}^{u}_{\bar{m}})^{H}]&=&\frac{P_{p}(\beta^{u})^{2}}{P_{p}\beta^{u}+\sigma^{2}_{\eta}}\mathbf{I}_{N},\  \text{and} \label{eq:Cov_g}\\
\mathbb{E}[\mathbf{e}^{u}_{\bar{m}}(\mathbf{e}^{u}_{\bar{m}})^{H}]&=&\frac{\beta^{u}\sigma^{2}_{\eta}}{P_{p}\beta^{u}+\sigma^{2}_{\eta}}\mathbf{I}_{N}.\label{eq:Cov_e}
\end{eqnarray}
\subsubsection{MRC Receiver}
Employing $\mathbf{g}^{u}_{\bar{m}}=\mathbf{\hat{g}}^{u}_{\bar{m}}+\mathbf{e}^{u}_{\bar{m}}$ in \eqref{eq:MIMO_FD}, the estimate of the OQAM symbol at the MRC receiver output for the $u$th user at the FT index $({\bar{m},\bar{k}})$ can be formulated as
\begin{eqnarray}\label{eq:SC_MRC_IMCSI2}
\hat{d}^{u}_{\bar{m},\bar{k}} = \Re\big\{(\mathbf{\hat{g}}^{u}_{\bar{m}})^{H}\mathbf{y}_{\bar{m},\bar{k}}\big\}=\norm{\mathbf{\hat{g}}^{u}_{\bar{m}}}^{2}d^{u}_{\bar{m},\bar{k}}+v^{u,\text{mrc}}_{\bar{m},\bar{k}},
\end{eqnarray}
 where the real noise-plus-interference term $v^{u,\text{mrc}}_{\bar{m},\bar{k}}$ is expressed as
\begin{align}
v^{u,\text{mrc}}_{\bar{m},\bar{k}}\iftoggle{SINGLE_COL}{=}{&=}\Re\bigg\{\sum_{j=1,j\neq u}^{U}(\mathbf{\hat{g}}^{u}_{\bar{m}})^{H}\mathbf{\hat{g}}^{j}_{\bar{m}}b^{j}_{\bar{m},\bar{k}}+\sum_{j=1}^{U}(\mathbf{\hat{g}}^{u}_{\bar{m}})^{H}\mathbf{e}^{j}_{\bar{m}}b^{j}_{\bar{m},\bar{k}}\iftoggle{SINGLE_COL}{}{\\
\nonumber &}+(\mathbf{\hat{g}}^{u}_{\bar{m}})^{H}\boldsymbol{\eta}_{\bar{m},\bar{k}}\bigg\}.
\end{align}
Exploiting \eqref{eq:Cov_e}, \eqref{eq:Prel3} and the statistical properties of the intrinsic interference from \eqref{eq:Var_Intr}, the variance of the term $v^{u,\text{mrc}}_{\bar{m},\bar{k}}$ can be formulated as
 \begin{eqnarray}\label{eq:SC_MRC_IMCSI3}
\text{Var}\big[v^{u,\text{mrc}}_{\bar{m},\bar{k}}\big]=P_{d}\sum_{j=1,j\neq u}^{U}\big|(\mathbf{\hat{g}}^{u}_{\bar{m}})^{H}\mathbf{\hat{g}}^{j}_{\bar{m}}\big|^{2}+\dfrac{\sigma^{2}_{\eta}}{2}\norm{\mathbf{\hat{g}}^{u}_{\bar{m}}}^{2}\iftoggle{SINGLE_COL}{}{\\
\nonumber} +P_{d}\sum_{j=1}^{U}\frac{\beta^{j}\sigma^{2}_{\eta}}{P_{p}\beta^{j}+\sigma^{2}_{\eta}}\norm{\mathbf{\hat{g}}^{u}_{\bar{m}}}^{2}.
\end{eqnarray}
Following the rules given in \eqref{eq:R2C}, the MRC estimate of the symbol after OQAM to QAM conversion becomes:
\begin{eqnarray}\label{eq:SC_MRC_IMCSI2}
 \hat{c}^{u}_{\bar{m},\bar{k}} =\norm{\mathbf{\hat{g}}^{u}_{\bar{m}}}^{2}c^{u}_{\bar{m},\bar{k}}+\tilde{v}^{u,\text{mrc}}_{\bar{m},\bar{k}},
\end{eqnarray}
where $c^{u}_{\bar{m},\bar{k}}={d}^{u}_{\bar{m},2\bar{k}}+j{d}^{u}_{\bar{m},2\bar{k}+1}$ and $\tilde{v}^{u,\text{mrc}}_{\bar{m},\bar{k}}=v^{u,\text{mrc}}_{\bar{m},2\bar{k}}+jv^{u,\text{mrc}}_{\bar{m},2\bar{k}+1}$ when $\bar{m}$ is even, and for odd $\bar{m}$, $c^{u}_{\bar{m},\bar{k}}={d}^{u}_{\bar{m},2\bar{k}+1}+j{d}^{u}_{\bar{m},2\bar{k}}$ and $\tilde{v}^{u,\text{mrc}}_{\bar{m},\bar{k}}=v^{u,\text{mrc}}_{\bar{m},2\bar{k}+1}+jv^{u,\text{mrc}}_{\bar{m},2\bar{k}}$. Since the interference-plus-noise terms $v^{u,\text{mrc}}_{\bar{m},2\bar{k}}$ and $v^{u,\text{mrc}}_{\bar{m},2\bar{k}+1}$ are zero-mean independent with equal variances, the term  $\tilde{v}^{u,\text{mrc}}_{\bar{m},\bar{k}}$ after OQAM to QAM conversion has a variance of $\text{Var}[\tilde{v}^{u,\text{mrc}}_{\bar{m},\bar{k}}]=2\text{Var}[v^{u,\text{mrc}}_{\bar{m},\bar{k}}]$. Thus, the SINR at the $\bar{m}$th subcarrier of the $u$th user with imperfect CSI can be expressed as
\begin{eqnarray}\label{eq:SC_MRC_IMCSI3}
\Upsilon^{u,\text{mrc}}_{\bar{m},\text{IP}}=\dfrac{2P_{d}\norm{\mathbf{\hat{g}}^{u}_{\bar{m}}}^{2}}{2P_{d}\Big(\sum_{j=1,j\neq u}^{U}\big|\tilde{\mathbf{{\text{g}}}}^{j}_{\bar{m}}\big|^{2}+\sum_{j=1}^{U}\frac{\beta^{j}\sigma^{2}_{\eta}}{P_{p}\beta^{j}+\sigma^{2}_{\eta}}\Big)+\sigma^{2}_{\eta}},
\end{eqnarray}
where the random variable $\tilde{\text{g}}^{j}_{\bar{m}}$ obeys $\tilde{\text{g}}^{j}_{\bar{m}}=(\mathbf{\hat{g}}^{u}_{\bar{m}})^{H}\mathbf{\hat{g}}^{j}_{\bar{m}}/\norm{\mathbf{\hat{g}}^{u}_{\bar{m}}}$. It follows from \eqref{eq:Prel1} and \eqref{eq:Cov_g} that $\tilde{\text{g}}^{j}_{\bar{m}}\sim \mathcal{CN}\big(0,\frac{P_{p}(\beta^{j})^{2}}{P_{p}\beta^{j}+\sigma^{2}_{\eta}}\big)$. Furthermore, conditioned on $\mathbf{\hat{g}}^{u}_{\bar{m}}$, the random variable $\tilde{\mathbf{{\text{g}}}}^{j}_{\bar{m}}$ is independent from $\mathbf{\hat{g}}^{u}_{\bar{m}}$. For a fixed $E^{u}$, let the power of the $u$th user be scaled as $2P_{d}=E^{u}/\sqrt{N}$, and $N$ grows large. Then, by exploiting \eqref{eq:Prel1} and the fact from \eqref{eq:Cov_g} that each element of the vector $\mathbf{\hat{g}}^{u}_{\bar{m}}$ has a variance of $\frac{P_{p}(\beta^{u})^{2}}{P_{p}\beta^{u}+\sigma^{2}_{\eta}}$, the SINR
$\Upsilon^{u,\text{mrc}}_{\bar{m},\text{IP}}\xrightarrow[]{N\rightarrow \infty}K(\beta^{u}E^{u})^{2}/\sigma^{4}_{\eta}$.
The ergodic achievable uplink rate at the $\bar{m}$th subcarrier of the $u$th user can now be obtained as
\begin{align}
\mathcal{R}^{u,\text{mrc}}_{\bar{m},\text{IP}}=\mathbb{E}\big[\text{log}_{2}(1+\Upsilon^{u,\text{mrc}}_{\bar{m},\text{IP}})\big].
\end{align}
Exploiting the convexity of $\text{log}(1+\frac{1}{x})$ and Jensen's inequality of $\mathbb{E}[f(x)]\geq f(\mathbb{E}[x])$, the lower bound on the achievable uplink rate is obtained as $\mathcal{R}^{u,\text{mrc}}_{\bar{m},\text{IP}}\geq \tilde{\mathcal{R}}^{u,\text{mrc}}_{\bar{m},\text{IP}}=\text{log}_{2}\big(1+\big(\mathbb{E}\big[1/\Upsilon^{u,\text{mrc}}_{\bar{m},\text{IP}}\big]\big)^{-1}\big)$. The term $\mathbb{E}\big[1/\Upsilon^{u,\text{mrc}}_{\bar{m},\text{IP}}\big]$ can be evaluated as
\begin{align}
\iftoggle{SINGLE_COL}{}{\nonumber} \mathbb{E}\bigg[\dfrac{1}{\Upsilon^{u,\text{mrc}}_{\bar{m},\text{IP}}}\bigg]&=\bigg(\sum_{j=1,j\neq u}^{U}\mathbb{E}\big[\big|\tilde{\text{g}}^{j}_{\bar{m}}\big|^{2}\big]+\sum_{j=1}^{U}\frac{\beta^{j}\sigma^{2}_{\eta}}{P_{p}\beta^{j}+\sigma^{2}_{\eta}}\iftoggle{SINGLE_COL}{}{\\&}+\dfrac{\sigma^{2}_{\eta}}{2P_{d}}\bigg)\mathbb{E}\bigg[\dfrac{1}{\norm{\mathbf{\hat{g}}^{u}_{\bar{m}}}^{2}}\bigg].
\end{align}
The identity $\mathbb{E}[\text{Tr}(\boldsymbol{W}^{-1})]=k/(k-m)$ for an $m\times m$ central complex Wishart distributed matrix $\boldsymbol{W}$ with $k$ ($k>m$) degree of freedom \cite{TulinoV04} yields $\mathbb{E}\big[{1}/{\norm{\mathbf{\hat{g}}^{u}_{\bar{m}}}^{2}}\big]=\frac{(\beta^{u}P_{p}+\sigma^{2}_{\eta})}{P_{p}(\beta^{u})^{2}(N-1)}$ for $N\geq2$. Thus, the achievable uplink rate of the MRC receiver is lower bounded as
\begin{align}\label{eq:SC_MRC_IMCSI5}
\iftoggle{SINGLE_COL}{}{\nonumber &} \mathcal{R}^{u,\text{mrc}}_{\bar{m},\text{IP}}\geq \tilde{\mathcal{R}}^{u,\text{mrc}}_{\bar{m},\text{IP}}=\iftoggle{SINGLE_COL}{}{\\&}\text{log}_{2}\left(1+\dfrac{P_{p}(N-1)(\beta^{u})^{2}}{(P_{p}\beta^{u}+\sigma^{2}_{\eta})\bigg(\sum_{j=1,j\neq u}^{U}\beta^{j}+\dfrac{\sigma^{2}_{\eta}}{2P_{d}}\bigg)+\beta^{u}\sigma^{2}_{\eta}}\right).
\end{align}
By setting $2P_{d}=E^{u}/\sqrt{N}$ for a fixed $E^{u}$, and $N\rightarrow \infty$, $\tilde{\mathcal{R}}^{u,\text{mrc}}_{\bar{m},\text{IP}}\rightarrow\text{log}_{2}\left(1+{K (E^{u}\beta^{u})^{2}}/{\sigma^{4}_{\eta}}\right)$.
\subsubsection{ZF Receiver}
Employing ZF combining in \eqref{eq:MIMO_FD}, the estimate of the OQAM symbol vector at the FT index $(\bar{m},\bar{k})$ in the presence of imperfect CSI can be formulated as
\begin{eqnarray}\label{eq:SC_ZF_IMCSI1}
\nonumber \mathbf{\hat{d}}_{\bar{m},\bar{k}} = \Re\big\{\mathbf{\hat{G}}^{\dagger}_{\bar{m}}\mathbf{y}_{\bar{m},\bar{k}}\big\}=\mathbf{d}_{\bar{m},\bar{k}}+\mathbf{v}^{\text{zf}}_{\bar{m},\bar{k}},
\end{eqnarray}
where $\mathbf{\hat{G}}^{\dagger}_{\bar{m}}=\big(\mathbf{G}_{\bar{m}}^{H}\mathbf{G}_{\bar{m}}\big)^{-1}\mathbf{G}_{\bar{m}}^{H}$ and $\mathbf{v}^{\text{zf}}_{\bar{m},\bar{k}}=\Re\big\{\mathbf{\hat{G}}^{\dagger}_{\bar{m}}\sum_{j=1}^{U}\mathbf{e}^{j}_{\bar{m}}b^{j}_{\bar{m},\bar{k}}+\mathbf{\hat{G}}^{\dagger}_{\bar{m}}\boldsymbol{\eta}_{\bar{m},\bar{k}}\big\}$ is the noise-plus-interference vector at the output of the ZF receiver. Using \eqref{eq:Prel3}, \eqref{eq:Cov_e} and the statistical properties of the intrinsic interference from \eqref{eq:Var_Intr}, the covariance matrix of the vector $\mathbf{v}^{\text{zf}}_{\bar{m},\bar{k}}$ is
 \begin{eqnarray}\label{eq:SC_ZF_IMCSI3_1}
\nonumber \mathbb{E}\big[\mathbf{v}^{\text{zf}}_{\bar{m},\bar{k}}(\mathbf{v}^{\text{zf}}_{\bar{m},\bar{k}})^{H}\big]={\bigg(\sum_{j=1}^{U}\frac{P_{d}\beta^{j}\sigma^{2}_{\eta}}{P_{p}\beta^{j}+\sigma^{2}_{\eta}}+\dfrac{\sigma^{2}_{\eta}}{2}\bigg)\Big(\mathbf{\hat{G}}^{H}_{\bar{m}}\mathbf{\hat{G}}_{\bar{m}}\Big)^{-1}}.
\end{eqnarray}
Using \eqref{eq:R2C}, the ZF estimate of the QAM symbol vector $\mathbf{c}_{\bar{m},\bar{k}}$ can now be computed as
\begin{eqnarray}\label{eq:SC_ZF_IMCSI2}
\mathbf{\hat{c}}_{\bar{m},\bar{k}} =\mathbf{c}_{\bar{m},\bar{k}}+\mathbf{\tilde{v}}^{\text{zf}}_{\bar{m},\bar{k}}.
\end{eqnarray}
Using the fact that $E\big[\mathbf{\tilde{v}}^{\text{zf}}_{\bar{m},\bar{k}}(\mathbf{\tilde{v}}^{\text{zf}}_{\bar{m},\bar{k}})^{H}\big]=2E\big[\mathbf{v}^{\text{zf}}_{\bar{m},\bar{k}}(\mathbf{v}^{\text{zf}}_{\bar{m},\bar{k}})^{H}\big]$, the SINR at the $\bar{m}$th subcarrier of the $u$th user can be derived as
 \begin{eqnarray}\label{eq:SC_ZF_IMCSI3}
 \Upsilon^{u,\text{zf}}_{\bar{m},\text{IP}}=\dfrac{2P_{d}}{\Big(2P_{d}\sum_{j=1}^{U}\frac{\beta^{j}\sigma^{2}_{\eta}}{P_{p}\beta^{j}+\sigma^{2}_{\eta}}+\sigma^{2}_{\eta}\Big)\Big[\big(\mathbf{\hat{G}}^{H}_{\bar{m}}\mathbf{\hat{G}}_{\bar{m}}\big)^{-1}\Big]_{u,u}},
\end{eqnarray}
where $\big[\big(\mathbf{\hat{G}}^{H}_{\bar{m}}\mathbf{\hat{G}}_{\bar{m}}\big)^{-1}\big]_{u,u}$ denotes the $u$th diagonal element of the matrix $\big(\mathbf{\hat{G}}^{H}_{\bar{m}}\mathbf{\hat{G}}_{\bar{m}}\big)^{-1}$. By choosing $2P_{d}=E^{u}/\sqrt{N}$ and using \eqref{eq:Prel1}, as $N\rightarrow\infty$, it follows that $\Upsilon^{u,\text{zf}}_{\bar{m},\text{IP}}\rightarrow K(\beta^{u}E^{u})^{2}/\sigma^{4}_{\eta}$. Consequently, the achievable uplink rate for the $u$th user becomes:
\begin{align}
\mathcal{R}^{u,\text{zf}}_{\bar{m},\text{IP}}=\mathbb{E}\big[\text{log}_{2}(1+\Upsilon^{u,\text{zf}}_{\bar{m},\text{IP}})\big]\rightarrow \text{log}_{2}\left(1+\dfrac{K(\beta^{u}E^{u})^{2}}{\sigma^{4}_{\eta}}\right).
\end{align}
Upon employing \eqref{eq:Cov_g}, it follows from\cite{TulinoV04}  that $\mathbb{E}\big[\big\{\big(\mathbf{\hat{G}}^{H}_{\bar{m}}\mathbf{\hat{G}}_{\bar{m}}\big)^{-1}\big\}_{u,u}\big]=\frac{P_{p}\beta^{u}+\sigma^{2}_{\eta}}{P_{p}(\beta^{u})^{2}(N-U)}$. Thus, the lower bound on the achievable uplink rate for the $u$th user is determined as
\begin{align}\label{eq:SC_ZF_IMCSI4}
\iftoggle{SINGLE_COL}{\nonumber}{\nonumber &} \mathcal{R}^{u,\text{zf}}_{\bar{m},\text{IP}}\geq \tilde{\mathcal{R}}^{u,\text{zf}}_{\bar{m},\text{IP}}=\text{log}_{2}\left(1+\dfrac{P_{p}(N-U)(\beta^{u})^{2}}{(P_{p}\beta^{u}+\sigma^{2}_{\eta})\Big(\sum_{j=1}^{U}\frac{\beta^{j}\sigma^{2}_{\eta}}{P_{p}\beta^{j}+\sigma^{2}_{\eta}}+\dfrac{\sigma^{2}_{\eta}}{2P_{d}}\Big)}\right).
\end{align}
Note that for $2P_{d}=E^{u}/\sqrt{N}$ and $N\rightarrow\infty$, $\tilde{\mathcal{R}}^{u,\text{zf}}_{\bar{m},\text{IP}}\rightarrow \mathcal{R}^{u,\text{zf}}_{\bar{m},\text{IP}}$. It is worth mentioning that the power scaling laws, similar to those of the OFDM-based MU massive MIMO systems in \cite{NgoLM13}, also hold for their FBMC counterparts.
\subsubsection{MMSE Receiver} Substituting $\mathbf{g}^{u}_{\bar{m}}=\mathbf{\hat{g}}^{u}_{\bar{m}}+\mathbf{e}^{u}_{\bar{m}}$ in \eqref{eq:MIMO_FD}, one obtains $\mathbf{y}_{\bar{m},\bar{k}} = \mathbf{\hat{g}}^{u}_{\bar{m}}b^{u}_{\bar{m},\bar{k}}+\sum_{j=1,j\neq u}^{U}\mathbf{\hat{g}}^{j}_{\bar{m}}b^{j}_{\bar{m},\bar{k}}+\sum_{j=1}^{U}\mathbf{e}^{j}_{\bar{m}}b^{j}_{\bar{m},\bar{k}}+\boldsymbol{\eta}_{\bar{m},\bar{k}}$. Let the noise-plus-error vector be $\boldsymbol{\tilde{\eta}}_{\bar{m},\bar{k}}=\sum_{j=1}^{U}\mathbf{e}^{j}_{\bar{m}}b^{j}_{\bar{m},\bar{k}}+\boldsymbol{\eta}_{\bar{m},\bar{k}}$. Using \eqref{eq:Cov_e} and the variance of the intrinsic interference derived in \eqref{eq:Var_Intr}, the covariance of the vector $\boldsymbol{\tilde{\eta}}_{\bar{m},\bar{k}}$ is determined as $\mathbb{E}[\boldsymbol{\tilde{\eta}}_{\bar{m},\bar{k}}\boldsymbol{\tilde{\eta}}_{\bar{m},\bar{k}}^{H}] = 2P_{d}\sum_{j=1}^{U}\frac{\beta^{j}\sigma^{2}_{\eta}}{P_{p}\beta^{j}+\sigma^{2}_{\eta}}\mathbf{I}_{N}+\sigma^{2}_{\eta}\mathbf{I}_{N}$. Thus, in the presence of the channel estimation error, the $u$th column $\mathbf{\hat{a}}^{u}_{\bar{m}}$ of the MMSE combiner matrix $\mathbf{\hat{A}}_{\bar{m}}$ is 
\begin{eqnarray}\label{eq:MMSE_IMCSI2}
\nonumber \mathbf{\hat{a}}^{u}_{\bar{m}}=\Big(\mathbf{\hat{R}}_{\bar{m}}^{-1}+\mathbf{\hat{g}}_{\bar{m}}^{u}\left(\mathbf{\hat{g}}_{\bar{m}}^{u}\right)^{H}\Big)^{-1}\mathbf{\hat{g}}_{\bar{m}}^{u}\stackrel{(a)}{=}\dfrac{\mathbf{\hat{R}}_{\bar{m}}\mathbf{\hat{g}}_{\bar{m}}^{u}}{1+\left(\mathbf{\hat{g}}_{\bar{m}}^{u}\right)^{H}\mathbf{R}_{\bar{m}}\mathbf{\hat{g}}_{\bar{m}}^{u}},
\end{eqnarray}
where the matrix $\mathbf{\hat{R}}_{\bar{m}}^{-1}$ obeys $\mathbf{\hat{R}}_{\bar{m}}^{-1}=\sum_{j=1,j\neq u}^{U}\mathbf{\hat{g}}_{\bar{m}}^{j}(\mathbf{\hat{g}}_{\bar{m}}^{j})^{H}+\sum_{j=1}^{U}\frac{\beta^{j}\sigma^{2}_{\eta}}{P_{p}\beta^{j}+\sigma^{2}_{\eta}}\mathbf{I}_{N}+\frac{\sigma^{2}_{\eta}}{2P_{d}}\mathbf{I}_{N}$. The equality (a) follows from the matrix inversion lemma $\big(\mathbf{A}+\mathbf{u}\mathbf{v}^{T}\big)^{-1}=\mathbf{A}^{-1}-\frac{\mathbf{A}^{-1}\mathbf{u}\mathbf{v}^{T}\mathbf{A}^{-1}}{1+\mathbf{v}^{T}\mathbf{A}^{-1}\mathbf{u}}$. The estimate of the OQAM symbol at the MMSE combiner output can now be determined as
\begin{eqnarray}\label{eq:MMSE_PCSI3}
 \hat{d}^{u}_{\bar{m},\bar{k}}=\Re\big\{\left(\mathbf{\hat{a}}_{\bar{m}}^{u}\right)^{H}\mathbf{y}_{\bar{m},\bar{k}}\big\}=\alpha^{u}_{\bar{m}}d^{u}_{\bar{m},\bar{k}}+v^{u,\text{mmse}}_{\bar{m},\bar{k}},
\end{eqnarray}
where  $v^{u,\text{mmse}}_{\bar{m},\bar{k}}=\Re\big\{\sum_{j=1,j\neq u}^{U}\left(\mathbf{\hat{a}}_{\bar{m}}^{u}\right)^{H}\mathbf{\hat{g}}^{j}_{\bar{m}}b^{j}_{\bar{m},\bar{k}}+\sum_{j=1}^{U}\left(\mathbf{\hat{a}}_{\bar{m}}^{u}\right)^{H}\mathbf{e}^{j}_{\bar{m}}b^{j}_{\bar{m},\bar{k}}+\left(\mathbf{\hat{a}}_{\bar{m}}^{u}\right)^{H}\boldsymbol{\eta}_{\bar{m},\bar{k}}\big\}$ is the noise-plus-interference term  and the scalar $\alpha^{u}_{\bar{m}}=\left(\mathbf{\hat{a}}_{\bar{m}}^{u}\right)^{H}\mathbf{\hat{g}}_{\bar{m}}^{u}$. Since the matrix $\mathbf{\hat{R}}_{\bar{m}}$ is positive definite in nature, $\alpha^{u}_{\bar{m}}$ is a real and positive quantity.    Using \eqref{eq:Prel3}, \eqref{eq:Cov_e} and the property of the intrinsic interference  from \eqref{eq:Var_Intr}, the variance of the term $v^{u,\text{mmse}}_{\bar{m},\bar{k}}$ can be expressed as
\begin{align}
\iftoggle{SINGLE_COL}{\nonumber}{\nonumber &} \text{Var}\big[v^{u,\text{mmse}}_{\bar{m},\bar{k}}\big]\iftoggle{SINGLE_COL}{&}{} =P_{d}\sum_{j=1,j\neq u}^{U}\big|\left(\mathbf{\hat{a}}_{\bar{m}}^{u}\right)^{H}\mathbf{\hat{g}}_{\bar{m}}^{j}\big|^{2}+\frac{\sigma^{2}_{\eta}}{2}\norm{\mathbf{\hat{a}}_{\bar{m}}^{u}}^{2}\\
&+P_{d}\sum_{j=1}^{U}\frac{\beta^{j}\sigma^{2}_{\eta}}{P_{p}\beta^{j}+\sigma^{2}_{\eta}}\norm{\mathbf{\hat{a}}_{\bar{m}}^{u}}^{2}=P_{d}\left(\mathbf{\hat{a}}_{\bar{m}}^{u}\right)^{H}\mathbf{\hat{R}}^{-1}_{\bar{m}}\mathbf{\hat{a}}_{\bar{m}}^{u}.
\end{align}
Employing the rules given in \eqref{eq:R2C}, the MMSE estimate of the QAM symbol is
\begin{eqnarray}\label{eq:MMSE_PCSI6}
 \hat{c}^{u}_{\bar{m},\bar{k}}=\alpha^{u}_{\bar{m}}c^{u}_{\bar{m},\bar{k}}+\tilde{v}^{u,\text{mmse}}_{\bar{m},\bar{k}}.
\end{eqnarray}
Using the fact that the term $\tilde{v}^{u,\text{mmse}}_{\bar{m},\bar{k}}$ has a variance of $\text{Var}[\tilde{v}^{u,\text{mmse}}_{\bar{m},\bar{k}}]=2\text{Var}[v^{u,\text{mmse}}_{\bar{m},\bar{k}}]$, the SINR for the $u$th user at the MMSE combiner output becomes:
\begin{eqnarray}\label{eq:MMSE_PCSI7}
 \Upsilon^{u,\text{mmse}}_{\bar{m},\text{IP}}=\dfrac{(\alpha^{u}_{\bar{m}})^{2}}{\left(\mathbf{\hat{a}}_{\bar{m}}^{u}\right)^{H}\mathbf{\hat{R}}^{-1}_{\bar{m}}\mathbf{\hat{a}}_{\bar{m}}^{u}}\leq \left(\mathbf{\hat{a}}_{\bar{m}}^{u}\right)^{H}\mathbf{\hat{R}}_{\bar{m}}\mathbf{\hat{a}}_{\bar{m}}^{u}.
\end{eqnarray}
The achievable ergodic uplink rate at the $\bar{m}$th subcarrier of the $u$th user is
 $\mathcal{R}^{u,\text{mmse}}_{\bar{m},\text{IP}}=\mathbb{E}\big[\text{log}_{2}\big(1+\left(\mathbf{\hat{a}}_{\bar{m}}^{u}\right)^{H}\mathbf{\hat{R}}_{\bar{m}}\mathbf{\hat{a}}_{\bar{m}}^{u}\big)\big]$. Using the identity $1+\left(\mathbf{\hat{a}}_{\bar{m}}^{u}\right)^{H}\mathbf{\hat{R}}_{\bar{m}}\mathbf{\hat{a}}_{\bar{m}}^{u}=1/\big[(\mathbf{I}_{U}+c_{0}\mathbf{\hat{G}}^{H}_{\bar{m}}\mathbf{\hat{G}}_{\bar{m}})^{-1}\big]_{u,u}$ \cite{NgoLM13}, one obtains
\begin{eqnarray}\label{eq:MMSE_IMCSI3}
\mathcal{R}^{u,\text{mmse}}_{\bar{m},\text{IP}}=\mathbb{E}\bigg[\text{log}_{2}\bigg(\dfrac{1}{\big[(\mathbf{I}_{U}+c_{0}\mathbf{\hat{G}}_{\bar{m}}^{H}\mathbf{\hat{G}}_{\bar{m}})^{-1}\big]_{u,u}}\bigg)\bigg],
\end{eqnarray}
where the constant $c_{0}=\big(\sum_{j=1}^{U}\frac{\beta^{j}\sigma^{2}_{\eta}}{P_{p}\beta^{j}+\sigma^{2}_{\eta}}+\frac{\sigma^{2}_{\eta}}{2P_{d}}\big)^{-1}$. The uplink rate is lower bounded as
$\mathcal{R}^{u,\text{mmse}}_{\bar{m},\text{IP}}\geq \mathcal{\tilde{R}}^{u,\text{mmse}}_{\bar{m},\text{IP}}=\text{log}_{2}\big(1+(\hat{\pi}^{u}-1)\hat{\theta}^{u}\big)$,
where the parameters $\hat{\pi}^{u}=\frac{(N-U+1+(U-1)\hat{\mu})^{2}}{N-U+1+(U-1)\hat{\kappa}}$ and $\hat{\theta}^{u}=\frac{N-U+1+(U-1)\hat{\kappa}}{N-U+1+(U-1)\hat{\mu}}\frac{P_{p}(\beta^{u})^{2}}{c_{0}(P_{p}\beta^{u}+\sigma^{2}_{\eta})}$. The constants $\hat{\mu}$ and $\hat{\kappa}$ are computed using the rules in \cite[eq. (50)]{NgoLM13}.

\subsection{Perfect CSI}\label{SC_MU_FBMC_PCSI}
Using similar steps as in Section-\ref{SC_MU_FBMC_IMCSI}, the achievable uplink rate for the MRC, ZF and MMSE combining at the BS with perfect CSI can be determined as follows.
\subsubsection{MRC Receiver}
The SINR at $\bar{m}$th subcarrier of the $u$th user can be shown to be:
\begin{eqnarray}\label{eq:SINR_u_mrc}
\Upsilon^{u,\text{mrc}}_{\bar{m},\text{P}}= \dfrac{2P_{d}\norm{\mathbf{g}^{u}_{\bar{m}}}^{4}}{2P_{d}\sum_{i=1,i\neq u}^{U}\big\vert{(\mathbf{g}^{u}_{\bar{m}})^{H}\mathbf{g}^{i}_{\bar{m}}}\big\vert^{2}+\sigma^{2}_{\eta}\norm{\mathbf{g}^{u}_{\bar{m}}}^{2}}.
\end{eqnarray}
The  asymptotic SINR and uplink rate are determined~as $\Upsilon^{u,\text{mrc}}_{\bar{m},\text{P}}\big|_{2P_{d}=E^{u}/N}\xrightarrow[]{N\rightarrow \infty}{\beta^{u}E^{u}}/{\sigma^{2}_{\eta}}$ and $\mathcal{R}^{u,\text{mrc}}_{\bar{m},\text{P}}=\nonumber \mathbb{E}\big[\text{log}_{2}(1+\Upsilon^{u,\text{mrc}}_{\bar{m},\text{P}})\big]\xrightarrow[]{N\rightarrow \infty} \text{log}_{2}\big(1+\frac{\beta^{u}E^{u}}{\sigma^{2}_{\eta}}\big)$.
The achievable rate is lower-bounded as
\begin{eqnarray}\label{eq:Rate_LB}
\nonumber \mathcal{R}^{u,\text{mrc}}_{\bar{m},\text{P}}\geq \tilde{\mathcal{R}}^{u,\text{mrc}}_{\bar{m},\text{P}}=\text{log}_{2}\left(1+\dfrac{2P_{d}(N-1)\beta^{u}}{2P_{d}\sum_{i=1,i\neq u}^{U}\beta^{i}+\sigma^{2}_{\eta}}\right).
\end{eqnarray}
It can also be verified that for $2P_{d}=E^{u}/N$ and $N\rightarrow\infty$, the lower-bound $\tilde{\mathcal{R}}^{u,\text{mrc}}_{\bar{m},\text{P}}\rightarrow \mathcal{R}^{u,\text{mrc}}_{\bar{m},\text{P}}$.
\subsubsection{ZF Receiver}
The SINR at the $\bar{m}$th subcarrier of the $u$th user is obtained as
\begin{eqnarray}\label{eq:SINR_zf}
\Upsilon^{u,\text{zf}}_{\bar{m},\text{P}}= \dfrac{2P_{d}}{\sigma^{2}_{\eta}\Big\{\big(\mathbf{G}_{\bar{m}}^{H}\mathbf{G}_{\bar{m}}\big)^{-1}\Big\}_{u,u}}.
\end{eqnarray}
 The corresponding lower-bound on the achievable rate $\mathcal{R}^{u,\text{zf}}_{\bar{m},\text{P}}=\nonumber \mathbb{E}\big[\text{log}_{2}(1+\Upsilon^{u,\text{zf}}_{\bar{m},\text{P}})\big]$ is
\begin{eqnarray}\label{eq:Rate_zf_LB1}
 \mathcal{R}^{u,\text{zf}}_{\bar{m},\text{P}}\geq \tilde{\mathcal{R}}^{u,\text{zf}}_{\bar{m},\text{P}}=\text{log}_{2}\left(1+\dfrac{2P_{d}\beta^{u}(N-U)}{\sigma^{2}_{\eta}}\right).
\end{eqnarray}
Setting $2P_{d}=E^{u}/N$,  as $N$ grows large, we have $\tilde{\mathcal{R}}^{u,\text{zf}}_{\bar{m},\text{P}}\xrightarrow[]{N\rightarrow \infty}\text{log}_{2}\left(1+{E^{u}\beta^{u}}/{\sigma^{2}_{\eta}}\right). $
\subsubsection{MMSE Receiver}
Similarly, for the MMSE receiver, the achievable ergodic uplink rate is
\begin{eqnarray}\label{eq:MMSE_PCSI8}
\mathcal{R}^{u,\text{mmse}}_{\bar{m},\text{P}}=\mathbb{E}\bigg[\text{log}_{2}\bigg(\dfrac{1}{\big[(\mathbf{I}_{U}+\frac{2P_{d}}{\sigma^{2}_{\eta}}\mathbf{G}^{H}_{\bar{m}}\mathbf{G}_{\bar{m}})^{-1}\big]_{u,u}}\bigg)\bigg].
\end{eqnarray}
The achievable uplink rate is lower-bounded as
$\mathcal{R}^{u,\text{mmse}}_{\bar{m},\text{P}}\geq \mathcal{\tilde{R}}^{u,\text{mmse}}_{\bar{m},\text{P}}=\text{log}_{2}\big(1+(\pi^{u}-1)\theta^{u}\big)$,
where the parameters obey $\pi^{u}=\frac{(N-U+1+(U-1)\mu)^{2}}{N-U+1+(U-1)\kappa}$ and $\theta^{u}=\frac{N-U+1+(U-1)\kappa}{N-U+1+(U-1)\mu}\frac{2P_{d}}{\sigma^{2}_{\eta}}\beta^{u}$. The constants $\mu$ and $\kappa$ are computed using the rules given in \cite[eq. (28)]{NgoLM13}.

\section{Multi-Cell MU Massive MIMO-FBMC System}\label{MC_MU_MMIMO}
Let us now consider the uplink of a multi-cell MU MIMO-FBMC system with $N_{c}$ cells sharing the same frequency band. Each of the cells consists of a single BS equipped with $N$ antennas and $U$ single-antenna users. From \eqref{eq:MIMO_FD}, the $N\times 1$ receive vector at subcarrier index $\bar{m}$ and symbol time index $\bar{k}$ at the $n$th BS can be expressed as
\begin{eqnarray}\label{eq:MC_R1}
 \mathbf{y}_{\bar{m},\bar{k},n}= \sum_{i=1}^{N_{c}}\mathbf{G}_{\bar{m},n,i}\mathbf{b}_{\bar{m},\bar{k},i}+\boldsymbol{\eta}_{\bar{m},\bar{k},n},
\end{eqnarray}
where $\mathbf{G}_{\bar{m},n,i}=[\mathbf{g}^{1}_{\bar{m},n,i},\mathbf{g}^{2}_{\bar{m},n,i},\ldots,\mathbf{g}^{U}_{\bar{m},n,i}]\in \mathbb{C}^{N\times U}$ denotes the CFR matrix at the $\bar{m}$th subcarrier between the $n$th BS and the $U$ users in the $i$th cell, $\mathbf{b}_{\bar{m},\bar{k},i}\in \mathbb{C}^{U\times 1}$ is the virtual symbol vector of the $U$ users in the $i$th cell and $\boldsymbol{\eta}_{\bar{m},\bar{k},n}\in \mathbb{C}^{N\times 1}$ is the noise vector at the $n$th BS. Similar to the single-cell scenario in \eqref{eq:G_Mtx}, the CFR matrix $\mathbf{G}_{\bar{m},n,i}$ for the multi-cell scenario is modelled as
\begin{equation}
\mathbf{G}_{\bar{m},n,i}=\mathbf{H}_{\bar{m},n,i}\mathbf{D}^{1/2}_{n,i},
\end{equation}
where the matrix $\mathbf{H}_{\bar{m},n,i}$ comprises the fading coefficients at the $\bar{m}$th subcarrier between the $n$th BS station and the $U$ users in the $i$th cell. The $U\times U$ diagonal matrix $\mathbf{D}^{1/2}_{n,i}$ comprises the large-scale fading and the shadowing factors  between the $n$th BS station and $U$ users in the $i$th cell such that $[\mathbf{D}_{n,i}]_{(u,u)}=\beta^{u}_{n,i}$ for $i\neq n$ and $[\mathbf{D}_{n,n}]_{(u,u)}=\beta^{u}_{n,n}=1$.  The elements of the matrix $\mathbf{H}_{\bar{m},n,i}$ are modelled as i.i.d. $\mathcal{CN}(0,1)$.
\subsection{Perfect CSI}\label{MC_PCSI}
\subsubsection{MRC Receiver}
The OQAM symbol estimate at the output of the MRC receiver at the $n$th BS for the $u$th user at the FT index $(\bar{m},\bar{k})$ is
\begin{eqnarray}\label{eq:MRC_P1}
\nonumber \hat{d}^{u}_{\bar{m},\bar{k},n} = \Re\big\{(\mathbf{g}^{u}_{\bar{m},n,n})^{H}\mathbf{y}_{\bar{m},\bar{k},n}\big\}=\norm{\mathbf{g}^{u}_{\bar{m},n,n}}^{2}{d}^{u}_{\bar{m},\bar{k},n}+w^{u,\text{mrc}}_{\bar{m},\bar{k},n},
\end{eqnarray}
where ${d}^{u}_{\bar{m},\bar{k},n}=\Re\big\{{b}^{u}_{\bar{m},\bar{k},n}\big\}$ denotes the OQAM symbol transmitted by the $u$th user in the $n$th cell at the FT index $(\bar{m},\bar{k})$ and the noise-plus-interference term $w^{u}_{\bar{m},\bar{k},n}$ is expressed as
\begin{align}\label{eq:MRC_P2}
\nonumber & w^{u,\text{mrc}}_{\bar{m},\bar{k},n}=\Re\Bigg\{\sum_{i=1,i\neq n}^{N_{c}}\sum_{j=1}^{U}(\mathbf{g}^{u}_{\bar{m},n,n})^{H}\mathbf{g}^{j}_{\bar{m},n,i}b^{j}_{\bar{m},\bar{k},i}\iftoggle{SINGLE_COL}{}{\\
\nonumber &}+\sum_{j=1,j\neq u}^{U}(\mathbf{g}^{u}_{\bar{m},n,n})^{H}\mathbf{g}^{j}_{\bar{m},n,n}b^{j}_{\bar{m},\bar{k},n}+(\mathbf{g}^{u}_{\bar{m},n,n})^{H}\boldsymbol{\eta}_{\bar{m},\bar{k},n}\Bigg\}.
\end{align}
The first and second terms in the above equation represent the inter-cell-interference and intra-cell-interference, respectively. Using \eqref{eq:Prel3} and the statistical characteristics of the intrinsic interference from \eqref{eq:Var_Intr}, the variance of the noise-plus interference term $w^{u,\text{mrc}}_{\bar{m},\bar{k},n}$ can be formulated~as
\begin{align}
\iftoggle{SINGLE_COL}{\nonumber}{\nonumber &} \text{Var}[{w}^{u,\text{mrc}}_{\bar{m},\bar{k},n}]=P_{d}\sum_{i=1,i\neq n}^{N_{c}}\sum_{j=1}^{U}\big|(\mathbf{g}^{u}_{\bar{m},n,n})^{H}\mathbf{g}^{j}_{\bar{m},n,i}\big|^{2}\iftoggle{SINGLE_COL}{}{\\ &}+P_{d}\sum_{j=1,j\neq u}^{U}\big|(\mathbf{g}^{u}_{\bar{m},n,n})^{H}\mathbf{g}^{j}_{\bar{m},n,n}\big|^{2}+\dfrac{{\sigma^{2}_{\eta}}}{2}\norm{\mathbf{g}^{u}_{\bar{m},n,n}}^{2}.
\end{align}
The estimated QAM symbol after OQAM to QAM conversion~is
\begin{eqnarray}\label{eq:MRC_P3}
\hat{c}^{u}_{\bar{m},\bar{k},n}=\norm{\mathbf{g}^{u}_{\bar{m},n,n}}^{2}{c}^{u}_{\bar{m},\bar{k},n}+\tilde{w}^{u,\text{mrc}}_{\bar{m},\bar{k},n}.
\end{eqnarray}
Here ${c}^{u}_{\bar{m},\bar{k},n}={d}^{u}_{\bar{m},2\bar{k},n}+j{d}^{u}_{\bar{m},2\bar{k}+1,n}$ and $\tilde{w}^{u,\text{mrc}}_{\bar{m},\bar{k},n}={w}^{u,\text{mrc}}_{\bar{m},2\bar{k},n}+j{w}^{u,\text{mrc}}_{\bar{m},2\bar{k}+1,n}$ if subcarrier index $\bar{m}$ is even, and for odd $\bar{m}$, ${c}^{u}_{\bar{m},\bar{k},n}={d}^{u}_{\bar{m},2\bar{k}+1,n}+j{d}^{u}_{\bar{m},2\bar{k},n}$ and $\tilde{w}^{u,\text{mrc}}_{\bar{m},\bar{k},n}=w^{u,\text{mrc}}_{\bar{m},2\bar{k}+1,n}+jw^{u,\text{mrc}}_{\bar{m},2\bar{k},n}$. Since the terms $w^{u,\text{mrc}}_{\bar{m},2\bar{k},n}$ and $w^{u,\text{mrc}}_{\bar{m},2\bar{k}+1,n}$ are zero-mean independent with equal variances, we get $\text{Var}[\tilde{w}^{u,\text{mrc}}_{\bar{m},\bar{k},n}]=2\text{Var}[{w}^{u,\text{mrc}}_{\bar{m},\bar{k},n}]$. Using \eqref{eq:MRC_P3}, the SINR at the $n$th BS for the $u$th user is obtained as
\begin{align}
 &  {\Upsilon}^{u,\text{mrc}}_{\bar{m},n,\text{P}}=\dfrac{2P_{d}\big|\big|\mathbf{g}^{u}_{\bar{m},n,n}\big|\big|^{4}}{2\text{Var}[{w}^{u,\text{mrc}}_{\bar{m},\bar{k},n}]}.
\end{align}
It can be verified that by setting $2P_{d}=E^{u}/N$ and $N\rightarrow\infty$, we have $\Upsilon^{u,\text{mrc}}_{\bar{m},n,\text{P}}\rightarrow \beta^{u}_{n,n}E^{u}/\sigma^{2}_{\eta}$. Thus, similar to single-cell MU massive MIMO-FBMC systems, the power scaling law also holds in the case of multi-cell MU massive MIMO-FBMC systems. Next, the achievable uplink rate of $\mathcal{R}^{u,\text{mrc}}_{\bar{m},n,\text{P}}=\mathbb{E}\big[\text{log}_{2}\big(1+{\Upsilon}^{u,\text{mrc}}_{\bar{m},n,\text{P}}\big)\big]\xrightarrow{N\rightarrow\infty}\text{log}_{2}\big(1+\beta^{u}_{n,n}E^{u}/\sigma^{2}_{\eta}\big)$. Using the identity that $\mathbb{E}[1/||\mathbf{g}^{u}_{\bar{m},n,n}||^{2}]=1/[\beta^{u}_{n,n}(N-1)]$, the lower bound on the achievable uplink rate is
\begin{align}
\mathcal{{R}}^{u,\text{mrc}}_{\bar{m},n,\text{P}}\geq\tilde{\mathcal{R}}^{u,\text{mrc}}_{\bar{m},n,\text{P}}=\text{log}_{2}\left(1+\dfrac{2P_{d}(M-1)\beta^{u}_{n,n}}{ 2P_{d}\Big(\sum_{i=1,i\neq n}^{N_{c}}\sum_{j=1}^{U}\beta^{j}_{n,i}+\sum_{j=1,j\neq u}^{U}\beta^{j}_{n,n}\Big)+\sigma^{2}_{\eta}}\right).
\end{align}
\subsubsection{ZF Receiver} Following similar lines, the SINR can be expressed as
 \begin{align}
 {\Upsilon}^{u,\text{zf}}_{\bar{m},n,\text{P}}=\dfrac{2P_{d}}{\Big(2P_{d}\sum_{i=1,i\neq n}^{N_{c}}\sum_{j=1}^{U}\beta^{j}_{n,i}+\sigma^{2}_{\eta}\Big)\Big\{\Big(\mathbf{G}_{\bar{m},n,n}^{H}\mathbf{G}_{\bar{m},n,n}\Big)^{-1}\Big\}_{u,u}}.
\end{align}
The lower-bound on the achievable uplink rate is \begin{align}
\iftoggle{SINGLE_COL}{}{\nonumber }\tilde{\mathcal{R}}^{u,\text{zf}}_{\bar{m},n,\text{P}}&=\text{log}_{2}\left(1+\dfrac{2P_{d}(M-U)\beta^{u}_{n,n}}{ 2P_{d}\sum_{i=1,i\neq n}^{N_{c}}\sum_{j=1}^{U}\beta^{j}_{n,i}+\sigma^{2}_{\eta}}\right)\iftoggle{SINGLE_COL}{}{\\&}\xrightarrow[N\rightarrow \infty ]{2P_{d}=E^{u}/N}\text{log}_{2}\big(1+{\beta}^{u}_{n,n}E^{u}/\sigma^{2}_{\eta}\big).
\end{align}
\subsection{Imperfect CSI}\label{MC_IMCSI}
\subsubsection{Training-based linear MMSE Channel Estimation}
It is assumed that the users in each cell transmit the same set of training symbols according to the frame structure in Fig.~\ref{fig:frame}. By evaluating \eqref{eq:MC_R1} at the training symbol locations $k=2i$ for $0\leq i\leq K-1$ and stacking the resultant outputs, the received training symbol matrix $\mathbf{Y}_{\bar{m},n}=[\mathbf{y}_{\bar{m},0,n},\mathbf{y}_{\bar{m},2,n},\ldots,\mathbf{y}_{\bar{m},2(K-1),n}]\in \mathbb{C}^{N\times K}$ at the $n$th BS is expressed as
\begin{eqnarray}\label{eq:MC_CE1}
 \mathbf{Y}_{\bar{m},n}= \sum_{i=1}^{N_{c}}\mathbf{G}_{\bar{m},n,i}\mathbf{B}^{T}_{\bar{m}}+\mathbf{W}_{\bar{m},n},
\end{eqnarray}
where  $\mathbf{W}_{\bar{m},n}=[\boldsymbol{\eta}_{\bar{m},0,n},\boldsymbol{\eta}_{\bar{m},2,n},\ldots,\boldsymbol{\eta}_{\bar{m},2(K-1),n}]\in \mathbb{C}^{N\times K}$ is the corresponding noise matrix. Each element of the noise matrix $\mathbf{W}_{\bar{m},n}$ is distributed as $\mathcal{CN}(0,\sigma^{2}_{\eta})$. Upon exploiting the orthogonality among columns of the virtual training matrix $\mathbf{B}_{\bar{m}}$, the received training vector at the $n$th BS for the $u$th user in the $n$th cell can be evaluated as
\begin{align}\label{eq:MC_CE2}
\iftoggle{SINGLE_COL}{}{\nonumber}\mathbf{y}^{u}_{\bar{m},n,n}&=\mathbf{Y}_{\bar{m},n}\big(\mathbf{b}^{u}_{\bar{m}}\big)^{*}=P_{p}\mathbf{g}^{u}_{\bar{m},n,n}+\sum_{i=1,i\neq n}^{N_{c}}P_{p}\mathbf{g}^{u}_{\bar{m},n,i}\iftoggle{SINGLE_COL}{}{\\&}+\mathbf{W}_{\bar{m},n}\big(\mathbf{b}^{u}_{\bar{m}}\big)^{*}=P_{p}\mathbf{g}^{u}_{\bar{m},n,n}+\mathbf{w}^{u}_{\bar{m},n,n},
\end{align}
where the noise-plus-interference vector $\mathbf{w}^{u}_{\bar{m},n,n}=\sum_{i=1,i\neq n}^{N_{c}}P_{p}\mathbf{g}^{u}_{\bar{m},n,i}+\mathbf{W}_{\bar{m},n}\big(\mathbf{b}^{u}_{\bar{m}}\big)^{*}$. Note that the term $\sum_{i=1,i\neq n}^{N_{c}}P_{p}\mathbf{g}^{u}_{\bar{m},n,i}$ represents the inter-cell interference arising due to the pilot contamination. This term appears because of the pilot reuse among different cells.  Exploiting the second-order statistical properties of the intrinsic interference from \eqref{eq:Var_Intr}, it can be readily verified that the noise vector $\mathbf{W}_{\bar{m},n}\big(\mathbf{b}^{u}_{\bar{m}}\big)^{*}$ is distributed as $\mathcal{CN}(0,P_{p}\sigma^{2}_{\eta}\mathbf{I}_{N})$. The covariance matrix of the vector $\mathbf{g}^{u}_{\bar{m},n,n}$ is $\mathbf{C}_{\mathbf{g}^{u}_{\bar{m},n,n}}=\mathbb{E}[\mathbf{g}^{u}_{\bar{m},n,n}(\mathbf{g}^{u}_{\bar{m},n,n})^{H}]=\mathbf{I}_{N}$. Furthermore, the covariance matrix $\mathbf{C}_{\mathbf{w}^{u}_{\bar{m},n,n}}$ of the vector $\mathbf{w}^{u}_{\bar{m},n,n}$ can be determined as $\mathbf{C}_{\mathbf{w}^{u}_{\bar{m},n,n}}= (P_{p}^{2}(\gamma^{u}-1)+P_{p}\sigma^{2}_{\eta})\mathbf{I}_{N}$, where $\gamma^{u}=\sum_{i=1,i\neq n}^{N_{c}}\beta^{u}_{n,i}+1$. Upon using the above results, the MMSE estimate of the CFR vector  $\mathbf{g}^{u}_{\bar{m},n,n}$ at the $\bar{m}$th subcarrier between $n$th BS and $u$th user in the $n$th cell is now obtained as 
\begin{align}\label{eq:MC_CE3}
\iftoggle{SINGLE_COL}{}{\nonumber}\hat{\mathbf{g}}^{u}_{\bar{m},n,n}&= \dfrac{1}{P_{p}\gamma^{u}+\sigma^{2}_{\eta}}\mathbf{y}^{u}_{\bar{m},n,n}.
\end{align}
Upon using the expression for the variance of the intrinsic interference evaluated in \eqref{eq:Var_Intr}, the covariance matrices of the estimate $\hat{\mathbf{g}}^{u}_{\bar{m},n,n}$ and the error vector $\mathbf{e}^{u}_{\bar{m},n,n}=\mathbf{g}^{u}_{\bar{m},n,n}-\hat{\mathbf{g}}^{u}_{\bar{m},n,n}$ are 
\begin{align}
\mathbf{C}_{\hat{\mathbf{g}}^{u}_{\bar{m},n,n}}&=\mathbb{E}\big[\hat{\mathbf{g}}^{u}_{\bar{m},n,n}(\hat{\mathbf{g}}^{u}_{\bar{m},n,n})^{H}\big]= \dfrac{P_{p}}{P_{p}\gamma^{u}+\sigma^{2}_{\eta}}\mathbf{I}_{N},\label{eq:MC_CE4}\\
\mathbf{C}_{\mathbf{e}^{u}_{\bar{m},n,n}}&=\mathbb{E}\big[\mathbf{e}^{u}_{\bar{m},n,n}(\mathbf{e}^{u}_{\bar{m},n,n})^{H}\big]= \dfrac{P_{p}(\gamma^{u}-1)+\sigma^{2}_{\eta}}{P_{p}\gamma^{u}+\sigma^{2}_{\eta}}\mathbf{I}_{N}. \label{eq:MC_CE5}
\end{align}
Similar to (\ref{eq:MC_CE2}), the received training vector at the $n$th BS for the $u$th user in the $j$th cell is
\begin{align}\label{eq:MC_CE6}
\mathbf{y}^{u}_{\bar{m},n,j}= \mathbf{Y}_{\bar{m},n}\big(\mathbf{b}^{u}_{\bar{m}}\big)^{*}=P_{p}\mathbf{g}^{u}_{\bar{m},n,j}+\mathbf{w}^{u}_{\bar{m},n,j},
\end{align}
where the noise-plus-interference vector $\mathbf{w}^{u}_{\bar{m},n,j}$ at the $n$th BS for the $u$th user in the $j$th cell is expressed as $\mathbf{w}^{u}_{\bar{m},n,j}=P_{p}\mathbf{g}^{u}_{\bar{m},n,n}+\sum_{i=1,i\neq (j,n)}^{N_{c}}P_{p}\mathbf{g}^{u}_{\bar{m},n,i}+\mathbf{W}_{\bar{m},n}\big(\mathbf{b}^{u}_{\bar{m}}\big)^{*}$. Since $\mathbf{C}_{\mathbf{g}^{u}_{\bar{m},n,j}}=\beta^{u}_{n,j} \mathbf{I}_{N}$, it can be verified using \eqref{eq:Var_Intr} that  $\mathbf{C}_{\mathbf{w}^{u}_{\bar{m},n,j}}=(P^{2}_{p}\gamma^{u}-P^{2}_{p}\beta^{u}_{n,j}+P_{p}\sigma^{2}_{\eta})\mathbf{I}_{N}$. From \eqref{eq:MC_CE6}, the estimate of the CFR vector at the $\bar{m}$th subcarrier between the $n$th BS and the $u$th user in the $j$th cell is 
\begin{eqnarray}\label{eq:MC_CE7}
 \hat{\mathbf{g}}^{u}_{\bar{m},n,j}=\dfrac{\beta^{u}_{n,j}}{P_{p}\gamma^{u}+\sigma^{2}_{\eta}}\mathbf{y}^{u}_{\bar{m},n,j}=\beta^{u}_{n,j}\hat{\mathbf{g}}^{u}_{\bar{m},n,n},
\end{eqnarray}
where the last equality above follows from \eqref{eq:MC_CE3}. The covariance matrices of the vector $\hat{\mathbf{g}}^{u}_{\bar{m},n,j}$ and the corresponding estimation error vector $\mathbf{e}^{u}_{\bar{m},n,j}=\mathbf{g}^{u}_{\bar{m},n,j}-\hat{\mathbf{g}}^{u}_{\bar{m},n,j}$ are
\begin{align}\label{eq:MC_CE8}
\mathbf{C}_{\hat{\mathbf{g}}^{u}_{\bar{m},n,j}}&=(\beta^{u}_{n,j})^{2}\mathbb{E}[\hat{\mathbf{g}}^{u}_{\bar{m},n,n}(\hat{\mathbf{g}}^{u}_{\bar{m},n,n})^{H}]=\dfrac{P_{p}(\beta^{u}_{n,j})^{2}}{P_{p}\gamma^{u}+\sigma^{2}_{\eta}}\mathbf{I}_{N},\\
\mathbf{C}_{\mathbf{e}^{u}_{\bar{m},n,j}}&=\dfrac{\beta^{u}_{n,j} (P_{p}\gamma^{u}-P_{p}\beta^{u}_{n,j}+\sigma^{2}_{\eta})}{P_{p}\gamma^{u}+\sigma^{2}_{\eta}}\mathbf{I}_{N}.\label{eq:MC_CE9}
\end{align}
\subsubsection{MRC Receiver}
Employing $\mathbf{g}^{u}_{\bar{m},n,i}=\hat{\mathbf{g}}^{u}_{\bar{m},n,i}+\mathbf{e}^{u}_{\bar{m},n,i}$ in (\ref{eq:MC_R1}), the MRC estimate of the OQAM symbol at the $n$th BS for the $u$th user at the FT index $(\bar{m},\bar{k})$ can be formulated as
\begin{align}
\iftoggle{SINGLE_COL}{}{\nonumber} \hat{d}^{u}_{\bar{m},\bar{k},n}&= \Re\big\{(\mathbf{\hat{g}}^{u}_{\bar{m},n,n})^{H} \mathbf{y}_{\bar{m},\bar{k},n}\big\},\iftoggle{SINGLE_COL}{}{\\&}=\norm{\mathbf{\hat{g}}^{u}_{\bar{m},n,n}}^{2}{d}^{u}_{\bar{m},\bar{k},n}+v^{u,\text{mrc}}_{\bar{m},\bar{k},n},\label{eq:MC_R4}
\end{align}
where the noise-plus-interference term $v^{u,\text{mrc}}_{\bar{m},\bar{k},n}$ is expressed as
\begin{align}\label{eq:MC_R5}
\iftoggle{SINGLE_COL}{\nonumber}{\nonumber &}v^{u,\text{mrc}}_{\bar{m},\bar{k},n} \iftoggle{SINGLE_COL}{&}{}=\Re\Bigg\{\sum_{j=1,j\neq u}^{U}(\mathbf{\hat{g}}^{u}_{\bar{m},n,n})^{H}\mathbf{\hat{g}}^{j}_{\bar{m},n,n}{b}^{j}_{\bar{m},\bar{k},n}+\sum_{i=1,i\neq n}^{N_{c}}\sum_{j=1}^{U}(\mathbf{\hat{g}}^{u}_{\bar{m},n,n})^{H}\mathbf{e}^{j}_{\bar{m},n,i}{b}^{j}_{\bar{m},\bar{k},i}\\
&+\sum_{j=1}^{U}(\mathbf{\hat{g}}^{u}_{\bar{m},n,n})^{H}\mathbf{e}^{j}_{\bar{m},n,n}{b}^{j}_{\bar{m},\bar{k},n}+\sum_{i=1,i\neq n}^{N_{c}}\sum_{j=1}^{U}(\mathbf{\hat{g}}^{u}_{\bar{m},n,n})^{H}\hat{\mathbf{g}}^{j}_{\bar{m},n,i}{b}^{j}_{\bar{m},\bar{k},i}+(\mathbf{\hat{g}}^{u}_{\bar{m},n,n})^{H}\boldsymbol{\eta}_{\bar{m},\bar{k},n}\Bigg\}.
\end{align}
Using \eqref{eq:Prel3}, \eqref{eq:MC_CE5} and \eqref{eq:MC_CE9} along with the statistical properties of the intrinsic interference evaluated in \eqref{eq:Var_Intr}, the variance of the term $v^{u,\text{mrc}}_{\bar{m},\bar{k},n} $ above can be derived~as
\begin{align}\label{eq:App_B1}
\iftoggle{SINGLE_COL}{\nonumber}{\nonumber &}\text{Var}\big[v^{u,\text{mrc}}_{\bar{m},\bar{k},n} \big]& =P_{d}\sum_{i=1,i\neq n}^{N_{c}}\sum_{j=1,j\neq u}^{U}\big|(\mathbf{\hat{g}}^{u}_{\bar{m},n,n})^{H}\mathbf{\hat{g}}^{j}_{\bar{m},n,i}\big|^{2}+P_{d}\sum_{j=1,j\neq u}^{U}\big|(\mathbf{\hat{g}}^{u}_{\bar{m},n,n})^{H}\mathbf{\hat{g}}^{j}_{\bar{m},n,n}\big|^{2}\\
&+P_{d}\Bigg(\mu_{n}+\dfrac{\sigma^{2}_{\eta}}{2P_{d}}+\sum_{i=1,i\neq n}^{N_{c}}(\beta^{u}_{n,i})^{2}\norm{\hat{\mathbf{g}}^{u}_{\bar{m},n,n}}^{2}\Bigg)\norm{\hat{\mathbf{g}}^{u}_{\bar{m},n,n}}^{2},
\end{align}
where the quantity $\mu_{n}$ is defined as
\begin{eqnarray}\label{eq:MC_R7_1}
\mu_{n}&=&\sum_{i=1,i\neq n}^{N_{c}}\sum_{j=1}^{U}\dfrac{\beta^{j}_{n,i}(P_{p}\gamma^{j}-P_{p}\beta^{j}_{n,i}+\sigma^{2}_{\eta})}{P_{p}\gamma^{j}+\sigma^{2}_{\eta}}\iftoggle{SINGLE_COL}{+}{\\
\nonumber &+&}\sum_{j=1}^{U}\dfrac{P_{p}(\gamma^{j}-1)+\sigma^{2}_{\eta}}{P_{p}\gamma^{j}+\sigma^{2}_{\eta}}.
\end{eqnarray}
A detailed proof for the expression of the variance $\text{Var}\big[v^{u,\text{mrc}}_{\bar{m},\bar{k},n} \big]$ is given in Appendix-\ref{Signal_NPI}. Using the rules given in \eqref{eq:R2C}, the estimate of the QAM symbol  is obtained from \eqref{eq:MC_R4} as
\begin{eqnarray}\label{eq:MC_R6}
\hat{c}^{u}_{\bar{m},\bar{k},n}=\norm{\mathbf{\hat{g}}^{u}_{\bar{m},n,n}}^{2}{c}^{u}_{\bar{m},\bar{k},n}+\tilde{v}^{u,\text{mrc}}_{\bar{m},\bar{k},n}.
\end{eqnarray}
The variance of the noise-plus-interference $\tilde{v}^{u,\text{mrc}}_{\bar{m},\bar{k},n}$ is determined as $\text{Var}[\tilde{v}^{u,\text{mrc}}_{\bar{m},\bar{k},n}]=2\text{Var}\big[v^{u,\text{mrc}}_{\bar{m},\bar{k},n}\big]$. The SINR at the $\bar{m}$th subcarrier of the $u$th user at the $n$th BS  can now be expressed~as
\begin{eqnarray}\label{eq:MC_R8_1}
{\Upsilon}^{u,\text{mrc}}_{\bar{m},n,\text{IP}}= \dfrac{2P_{d}||\mathbf{\hat{g}}^{u}_{\bar{m},n,n}||^{4}}{2\text{Var}[{v}^{u,\text{mrc}}_{\bar{m},\bar{k},n}]}.
\end{eqnarray}
The ergodic uplink rate and the corresponding lower-bound at the $n$th BS for the $u$th user are
\begin{eqnarray}\label{eq:MC_R8}
\iftoggle{SINGLE_COL}{}{\nonumber} \mathcal{R}^{u,\text{mrc}}_{\bar{m},n,\text{IP}}&=&\mathbb{E}\big[\text{log}_{2}\big(1+{\Upsilon}^{u,\text{mrc}}_{\bar{m},n,\text{IP}}\big)\big]\geq\tilde{\mathcal{R}}^{u,\text{mrc}}_{\bar{m},n,\text{IP}}\iftoggle{SINGLE_COL}{=}{\\ &=&}\text{log}_{2}\big(1+\big\{\mathbb{E}\big[1/\Upsilon^{u,\text{mrc}}_{\bar{m},n,\text{IP}}\big]\big\}^{-1}\big).
\end{eqnarray}
The inverse SINR quantity $1/{\Upsilon}^{u,\text{mrc}}_{\bar{m},n,\text{IP}}$ is obtained as
\begin{align}
\iftoggle{SINGLE_COL}{\nonumber}{\nonumber}\dfrac{1}{{\Upsilon}^{u,\text{mrc}}_{\bar{m},n,\text{IP}}}&=\Bigg(\mu_{n}+\dfrac{\sigma^{2}_{\eta}}{2P_{d}}+\sum_{i=1,i\neq n}^{N_{c}}\sum_{j=1,j\neq u}^{U}\big|\tilde{\text{g}}^{j}_{\bar{m},n,i}\big|^{2}\iftoggle{SINGLE_COL}{}{\\&}+\sum_{j=1,j\neq u}^{U}\big|\tilde{\text{g}}^{j}_{\bar{m},n,n}\big|^{2}\Bigg)\dfrac{1}{||\mathbf{\hat{g}}^{u}_{\bar{m},n,n}||^{2}}+\sum_{i=1,i\neq n}^{N_{c}}(\beta^{u}_{n,i})^{2},
\end{align}
where $\tilde{\text{g}}^{j}_{\bar{m},n,n}=(\mathbf{\hat{g}}^{u}_{\bar{m},n,n})^{H}\mathbf{\hat{g}}^{j}_{\bar{m},n,n}/{||\mathbf{\hat{g}}^{u}_{\bar{m},n,n}||}$ and $\tilde{\text{g}}^{j}_{\bar{m},n,i}={(\mathbf{\hat{g}}^{u}_{\bar{m},n,n})^{H}\mathbf{\hat{g}}^{j}_{\bar{m},n,i}}/{||\mathbf{\hat{g}}^{u}_{\bar{m},n,n}||}$. Applying the result from \eqref{eq:Prel2}, and using \eqref{eq:MC_CE4} as well as \eqref{eq:MC_CE8}, it follows that $\tilde{\text{g}}^{j}_{\bar{m},n,n}$ and $\tilde{\text{g}}^{j}_{\bar{m},n,i}$ are zero mean Gaussian random variables with variances ${P_{p}}/({P_{p}\gamma^{j}+\sigma^{2}_{\eta}})$ and ${P_{p}(\beta^{j}_{n,i})^{2}}/{(P_{p}\gamma^{j}+\sigma^{2}_{\eta})}$, respectively, and are independent of  $\hat{\mathbf{g}}^{u}_{\bar{m},n,n}$.  Furthermore, since each element of the vector $\hat{\mathbf{g}}^{u}_{\bar{m},n,n}$ has a variance ${P_{p}}/{(P_{p}\gamma^{u}+\sigma^{2}_{\eta})}$, it follows from \cite{TulinoV04}  that $\mathbb{E}\big[{1}/{||\hat{\mathbf{g}}^{u}_{\bar{m},n,n}||^{2}}\big]={(P_{p}\gamma^{u}+\sigma^{2}_{\eta})}/{P_{p}(N-1)}$. Upon exploiting the above properties, one obtains
\begin{align}\label{eq:MC_R10}
\iftoggle{SINGLE_COL}{\nonumber}{\nonumber &}\mathbb{E}\bigg[\dfrac{1}{{\Upsilon}^{u,\text{mrc}}_{\bar{m},n,\text{IP}}}\bigg]\iftoggle{SINGLE_COL}{&}{}=\Bigg(\mu_{n}+\dfrac{\sigma^{2}_{\eta}}{2P_{d}}+\sum_{i=1,i\neq n}^{N_{c}}\sum_{j=1,j\neq u}^{U}\dfrac{P_{p}(\beta^{j}_{n,i})^{2}}{P_{p}\gamma^{j}+\sigma^{2}_{\eta}}\\
&+\sum_{j=1,j\neq u}^{U}\dfrac{P_{p}}{P_{p}\gamma^{j}+\sigma^{2}_{\eta}}\Bigg)\dfrac{(P_{p}\gamma^{u}+\sigma^{2}_{\eta})}{P_{p}(N-1)}+\sum_{i=1,i\neq n}^{N_{c}}(\beta^{u}_{n,i})^{2}.
\end{align}
On substituting $\big\{\mathbb{E}\big[1/{\Upsilon}^{u,\text{mrc}}_{\bar{m},n,\text{IP}}\big]\big\}^{-1}$ from above in (\ref{eq:MC_R8}), the lower bound on the achievable uplink rate at subcarrier $\bar{m}$ of the $u$th user at the $n$th BS for the multi-cell MU Massive MIMO-FBMC system in the presence of imperfect CSI can be determined as \iftoggle{SINGLE_COL}{}{given in \eqref{eq:MC_R11} (on the top of the next page).}
\iftoggle{SINGLE_COL}{}{\begin{figure*}}
\begin{align}\label{eq:MC_R11}
\iftoggle{SINGLE_COL}{\nonumber &}{} \tilde{\mathcal{R}}^{u,\text{mrc}}_{\bar{m},n,\text{IP}}\iftoggle{SINGLE_COL}{=\\ &}{=}\text{log}_{2}\left(1+\dfrac{2P_{d}P_{P}(N-1)}{({P_{p}\gamma^{u}+\sigma^{2}_{\eta}})\big( 2P_{d}U+2P_{d}\iftoggle{SINGLE_COL}{\displaystyle}{}\sum_{i=1,i\neq n}^{N_{c}}\sum_{j=1}^{U}{\beta^{j}_{n,i}}+{\sigma^{2}_{\eta}}\big)+2P_{d}P_{p}\big((N-2)\sum_{i=1,i\neq n}^{N_{c}}(\beta^{u}_{n,i})^{2}-1\big)}\right).
\end{align}
\iftoggle{SINGLE_COL}{}{\hrulefill
\end{figure*}}

\subsubsection{ZF Receiver}  The received OQAM symbol vector after ZF combining at the $n$th BS for the $U$ users in the $n$th cell can be formulated~as
\begin{align}\label{eq:ZFR3}
 \mathbf{\hat{d}}_{\bar{m},\bar{k},n}= \Re\big\{\mathbf{\hat{G}}^{\dagger}_{\bar{m},n,n}\mathbf{y}_{\bar{m},\bar{k},n}\big\}=\mathbf{d}_{\bar{m},\bar{k},n}+\mathbf{v}^{\text{zf}}_{\bar{m},\bar{k},n},
\end{align}
where $\mathbf{\hat{G}}^{\dagger}_{\bar{m},n,n}=(\mathbf{\hat{G}}^{H}_{\bar{m},n,n}\mathbf{\hat{G}}_{\bar{m},n,n})^{-1}\mathbf{\hat{G}}^{H}_{\bar{m},n,n}$   with $\mathbf{\hat{G}}_{\bar{m},n,n}=[\mathbf{\hat{g}}^{1}_{\bar{m},n,n},\mathbf{\hat{g}}^{2}_{\bar{m},n,n},\ldots,\mathbf{\hat{g}}^{U}_{\bar{m},n,n}]$, and the noise-plus-interference vector $\mathbf{v}^{\text{zf}}_{\bar{m},\bar{k},n}$ is expressed as
\begin{align}
\nonumber \mathbf{v}^{\text{zf}}_{\bar{m},\bar{k},n}= \Re\Bigg\{\mathbf{\hat{G}}^{\dagger}_{\bar{m},n,n}\Bigg(\sum_{i=1,i\neq n}^{N_{c}}\sum_{j=1}^{U}\mathbf{\hat{g}}^{j}_{\bar{m},n,i}{b}^{j}_{\bar{m},\bar{k},i}+\sum_{j=1}^{U}\mathbf{e}^{j}_{\bar{m},n,n}b^{j}_{\bar{m},\bar{k},n}+\sum_{i=1,i\neq n}^{N_{c}}\sum_{j=1}^{U}\mathbf{e}^{j}_{\bar{m},n,i}{b}^{j}_{\bar{m},\bar{k},i}+\boldsymbol{\eta}_{\bar{m},\bar{k},n}\Bigg)\Bigg\}.
\end{align}
By employing the results derived in  \eqref{eq:Var_Intr}, \eqref{eq:Prel3}, \eqref{eq:MC_CE4}, \eqref{eq:MC_CE5}, \eqref{eq:MC_CE8} and \eqref{eq:MC_CE9}, the covariance matrix of the noise-plus-interference term ${\mathbf{v}}^{\text{zf}}_{\bar{m},\bar{k},n}$ is determined as
\begin{align}\label{eq:ZFR5}
\iftoggle{SINGLE_COL}{\nonumber}{\nonumber &} \mathbb{E}\Big[{\mathbf{v}}^{\text{zf}}_{\bar{m},\bar{k},n}\big(\mathbf{v}^{\text{zf}}_{\bar{m},\bar{k},n}\big)^{H}\Big]\iftoggle{SINGLE_COL}{&}{} = P_{d}\Bigg(\mu_{n}+\sum_{i=1,i\neq n}^{N_{c}}\sum_{j=1,j\neq u}^{U}\dfrac{P_{p}(\beta^{j}_{n,i})^{2}}{P_{p}\gamma^{j}+\sigma^{2}_{\eta}}\\
&+\sum_{i=1,i\neq n}^{N}\dfrac{P_{p}(\beta^{u}_{n,i})^{2}}{P_{p}\gamma^{u}+\sigma^{2}_{\eta}}+\dfrac{\sigma^{2}_{\eta}}{2P_{d}}\Bigg)\left(\mathbf{\hat{G}}_{\bar{m},n,n}^{H}\mathbf{\hat{G}}_{\bar{m},n,n}\right)^{-1}.
\end{align}
After OQAM to QAM conversion, the ZF estimate of the QAM symbol vector becomes:
\begin{align}\label{eq:ZFR4}
 \mathbf{\hat{c}}_{\bar{m},\bar{k},n}= \mathbf{c}_{\bar{m},\bar{k},n}+\tilde{\mathbf{v}}^{\text{zf}}_{\bar{m},\bar{k},n},
\end{align}
where  $\tilde{\mathbf{v}}^{\text{zf}}_{\bar{m},\bar{k},n}=\mathbf{v}^{\text{zf}}_{\bar{m},2\bar{k},n}+j\mathbf{v}^{\text{zf}}_{\bar{m},2\bar{k}+1,n}$ when the subcarrier index $\bar{m}$ is even, and  $\tilde{\mathbf{v}}^{\text{zf}}_{\bar{m},\bar{k},n}=\mathbf{v}^{\text{zf}}_{\bar{m},2\bar{k}+1,n}+j\mathbf{v}^{\text{zf}}_{\bar{m},2\bar{k},n}$ otherwise. It can be verified that $\mathbb{E}\big[\tilde{\mathbf{v}}^{\text{zf}}_{\bar{m},\bar{k},n}\big(\tilde{\mathbf{v}}^{\text{zf}}_{\bar{m},\bar{k},n}\big)^{H}\big]=2\mathbb{E}\big[{\mathbf{v}}^{\text{zf}}_{\bar{m},\bar{k},n}\big(\mathbf{v}^{\text{zf}}_{\bar{m},\bar{k},n}\big)^{H}\big]$. The SINR at the $n$th BS for the $\bar{m}$th subcarrier of $u$th user can now be obtained from \eqref{eq:ZFR4} as
\begin{eqnarray}\label{eq:ZFR6}
{\Upsilon}^{u,\text{zf}}_{\bar{m},n,\text{IP}}= \dfrac{2P_{d}}{\Big\{\mathbb{E}\Big[\tilde{\mathbf{v}}^{\text{zf}}_{\bar{m},\bar{k},n}\big(\tilde{\mathbf{v}}^{\text{zf}}_{\bar{m},\bar{k},n}\big)^{H}\Big]\Big\}_{u,u}}.
\end{eqnarray}
Consequently, the achievable ergodic uplink rate for the $u$th user at the $\bar{m}th$ subcarrier is $\mathcal{R}^{u,\text{zf}}_{\bar{m},n,\text{IP}}=\mathbb{E}\big[\text{log}_{2}\big(1+{\Upsilon}^{u,\text{zf}}_{\bar{m},n,\text{IP}}\big)\big]$. Using \eqref{eq:MC_CE4}, it follows from \cite{TulinoV04} that $\mathbb{E}\big[\big\{\big(\mathbf{\hat{G}}^{H}_{\bar{m},n,n}\mathbf{\hat{G}}_{\bar{m},n,n}\big)^{-1}\big\}_{u,u}\big]=\frac{P_{p}\gamma^{u}+\sigma^{2}_{\eta}}{P_{p}(N-U)}$. Upon using the above properties, the lower-bound $\tilde{\mathcal{R}}^{u,\text{zf}}_{\bar{m},n,\text{IP}}$ on the achievable uplink rate at the $\bar{m}$th subcarrier of the $u$th user is given by: \iftoggle{SINGLE_COL}{}{in \eqref{eq:ZFR7} (on the top of the next page).}
\iftoggle{SINGLE_COL}{}{\begin{figure*}}
\begin{align} \label{eq:ZFR7}
 \tilde{\mathcal{R}}^{u,\text{zf}}_{\bar{m},n,\text{IP}}=\text{log}_{2}\left(1+\dfrac{2P_{d}P_{P}(N-U)}{({P_{p}\gamma^{u}+\sigma^{2}_{\eta}})\Big(2P_{d}U+2P_{d}\iftoggle{SINGLE_COL}{\displaystyle}{}\sum_{i=1,i\neq n}^{N_{c}}\sum_{j=1}^{U}{\beta^{j}_{n,i}}-\sum_{j=1,j\neq u}^{U}\dfrac{2P_{d}P_{p}}{P_{p}\gamma^{j}+\sigma^{2}_{\eta}}+{\sigma^{2}_{\eta}}\Big)-2P_{d}P_{p}}\right).
\end{align}
\iftoggle{SINGLE_COL}{}{\hrulefill \end{figure*}}

\section{Numerical Results}\label{Results}
\begin{table*}
    \centering
    \caption{Simulation Parameters}\label{tab:Par}
\begin{tabularx}{\linewidth}{|L|L|}
    \hline
\textbf{ Parameter} & \textbf{Specification}  \\ [0.5ex]
 \hline
Number of subcarrier ($M$)
& 128
\\
 \hline
Constellation  & $4$-QAM\\
\hline
Channel Model between a user and BS antenna pair
 & Complex Gaussian with $L=6$ equal power taps\\
 \hline
Symbol duration ($T$)
& $71.4\mu s$
\\
 \hline
Subcarrier spacing
 & $15k$Hz
\\
 \hline
Useful symbol duration
 & $66.7\mu s$
\\
\hline
Channel coherence time ($T_{c}$)
 & $1ms=196$ symbols \cite{NgoLM13}
\\
 \hline
 Number of users per cell ($U$)
 &$8$\\
\hline
 Number of training symbols per subcarrier ($K$)
 &$8$\\
 \hline
Prototype filter
 &IOTA with duration $4T\Rightarrow L_{p}=4M$\\
 \hline
 Noise variance ($\sigma_{\eta}^{2}$)
 &$1$\\
 \hline
\end{tabularx}
\end{table*}

Numerical examples are now presented to validate the various analytical results derived for the FBMC-based single- and multi-cell MU massive MIMO systems. The simulation parameters as summerized in Table-\ref{tab:Par} unless stated otherwise. 
The legend entries in the various plots are marked by the acronyms FBMC-ZF, FBMC-MRC, FBMC-MMSE, OFDM-ZF, OFDM-MRC, OFDM-MMSE that are self-explanatory. Furthermore, the perfect CSI, imperfect CSI and lower bound are denoted using the acronyms P-CSI, I-CSI and LB, respectively.

\subsection{Single-Cell Uplink Scenario}
The large-scale fading matrix $\mathbf{D}=\text{diag}[0.749\, 0.045\, 0.246\, 0.121\, 0.125\, 0.142\, 0.635\, 0.256]$ \cite{DaiD16} unless stated otherwise. The achievable uplink sum-rates at the $\bar{m}$th subcarrier for perfect and imperfect CSI at the BS are defined as $\sum_{u=1}^{U}\mathcal{R}^{u,\text{A}}_{\bar{m},\text{P}}$ and $\frac{T_{0}-K}{T_{0}}\sum_{u=1}^{U}\mathcal{R}^{u,\text{A}}_{\bar{m},\text{IP}}$, respectively with the corresponding expressions for the lower bounds given as $\sum_{u=1}^{U}\mathcal{\tilde{R}}^{u,\text{A}}_{\bar{m},\text{P}}$ and $\frac{T_{0}-K}{T_{0}}\sum_{u=1}^{U}\mathcal{\tilde{R}}^{u,\text{A}}_{\bar{m},\text{IP}}$, respectively \cite{NgoLM13}, where $T_{0}=196$, $\text{A}\in (\text{mrc, zf, mmse})$ and the quantities $\mathcal{R}^{u,\text{A}}_{\bar{m},\text{P}}$, $\mathcal{R}^{u,\text{A}}_{\bar{m},\text{IP}}$, $\mathcal{\tilde{R}}^{u,\text{A}}_{\bar{m},\text{P}}$ and $\mathcal{\tilde{R}}^{u,\text{A}}_{\bar{m},\text{IP}}$ are described in Section-\ref{SC_MU_FBMC}.
\begin{figure*}[t!]
\centering
\iftoggle{SINGLE_COL}{\begin{subfigure}[b]{.49\linewidth}}{\begin{subfigure}[b]{0.32\linewidth}}
   \includegraphics[width=\linewidth]{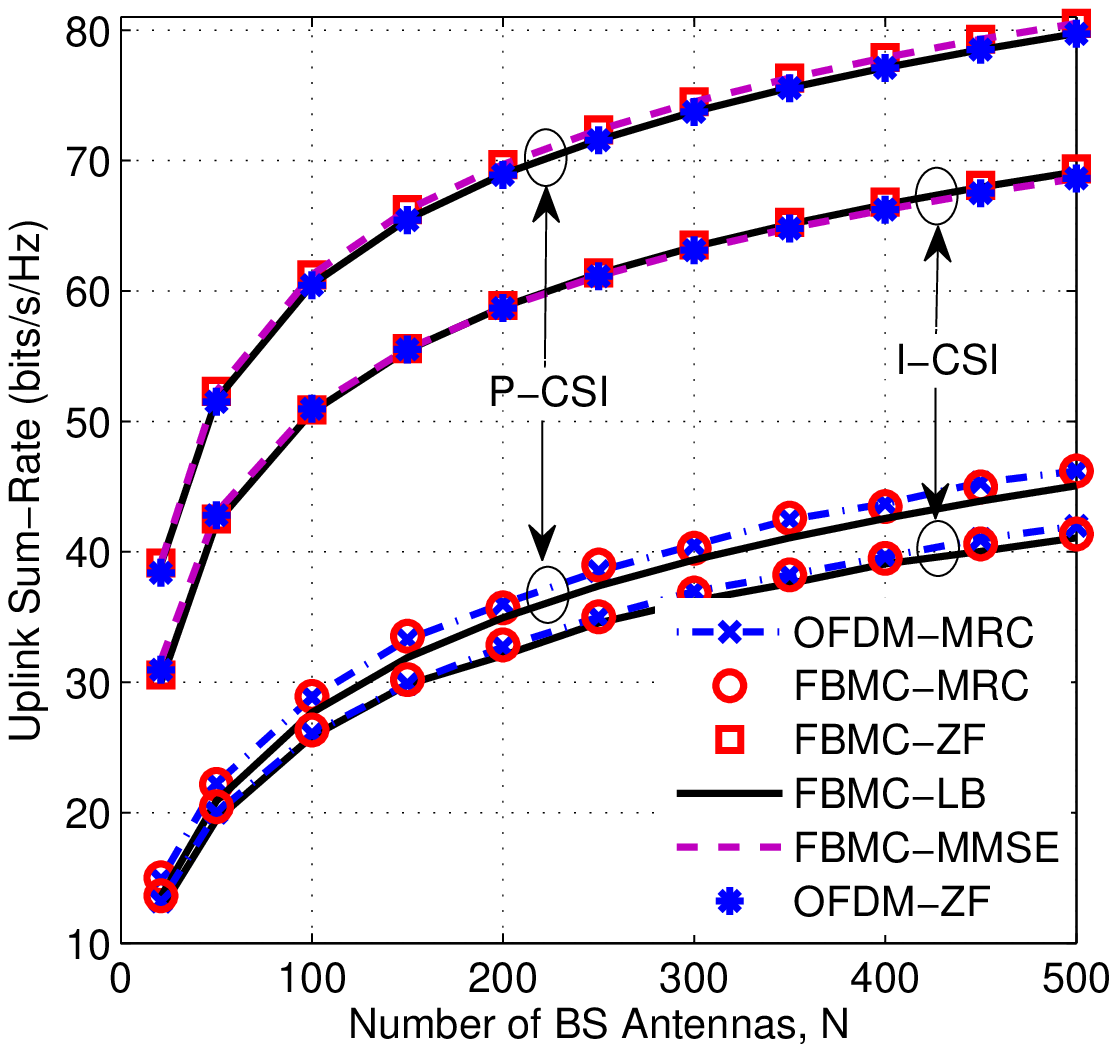}
   \caption{\small }
\label{fig:SC_AR_Vs_nRx_10dB}
\end{subfigure}
\iftoggle{SINGLE_COL}{\begin{subfigure}[b]{.49\linewidth}}{\begin{subfigure}[b]{0.32\linewidth}}
   \includegraphics[width=\linewidth]{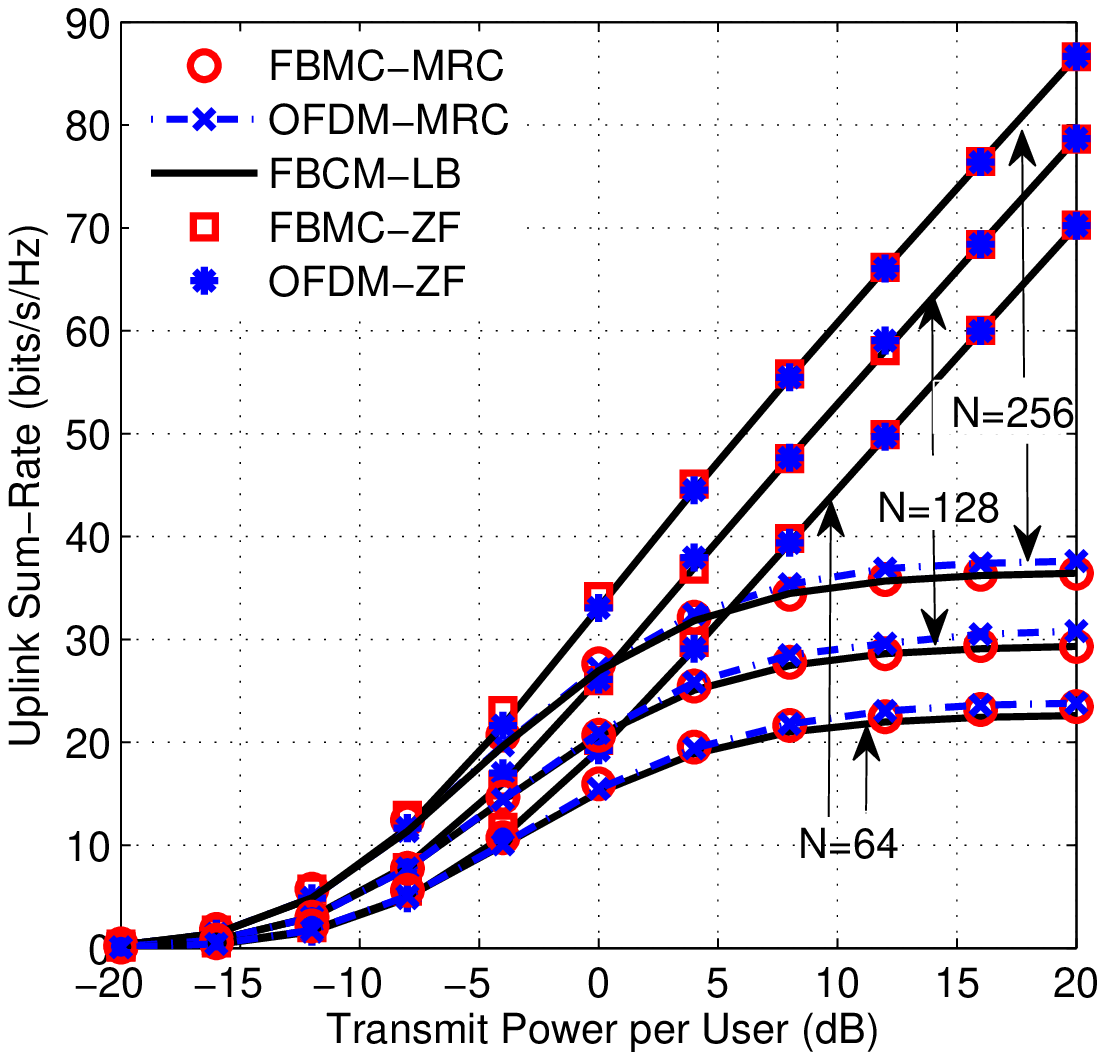}
   \caption{\small }
\label{fig:SC_AR_Vs_Pu}
\end{subfigure}
\caption{Uplink sum-rate versus  a) number of BS antennas for perfect and imperfect CSI with the transmit power per user $2P_{d}=10$ dB; and b) transmit power per user for different number of BS antennas with imperfect CSI.}
\end{figure*}

Fig.~\ref{fig:SC_AR_Vs_nRx_10dB} compares the achievable uplink sum-rates of the MRC, ZF and MMSE receivers to their corresponding lower bounds derived in Sections \ref{SC_MU_FBMC_IMCSI} and \ref{SC_MU_FBMC_PCSI} for scenarios with imperfect and perfect receive CSI, respectively. In other words, this study validates the analytical results derived in Sections \ref{SC_MU_FBMC_IMCSI} and \ref{SC_MU_FBMC_PCSI} for the FBMC-based massive MIMO system described in Section-\ref{System_Model} for the transmission over a quasi-static channel through a frequency flat response across all subcarriers. Clearly, for all the combiners, the simulated uplink sum-rates can be seen to precisely agree with their respective lower-bounds. It is also observed that the uplink sum-rate performance of the FBMC-based MU massive MIMO system relying on the MRC, ZF and MMSE receivers coincides with its counterparts in the CP-OFDM-based MU massive MIMO system.

Fig.~\ref{fig:SC_AR_Vs_Pu} shows the uplink sum-rate versus transmit power per user for the different number of BS antennas in the presence of imperfect CSI for both the FBMC and CP-OFDM-based MU massive MIMO systems using the ZF and MRC receivers. Similar to the previous figure, the uplink sum-rates of both the receivers can be seen to match their respective analytical lower-bounds derived in Section-\ref{SC_MU_FBMC_IMCSI}. Furthermore, both the combiners in the FBMC-based MU massive MIMO system  have a performance similar to the corresponding CP-OFDM system. Since the ZF combiner suppresses the multi-user-interference (MUI), it can be seen to outperform the MRC receiver in the high-power regime, where the effect of noise becomes negligible. On the other hand, the MRC receiver that maximises the received signal to noise ratio (SNR) suffers from the MUI. Thus, its performance  saturates when the transmit power per user increases. In the low power-regime, the effect of MUI decreases and noise begins to dominate. Therefore, the MRC receiver performs similar to that of the ZF receiver.
\begin{figure}[t!]
\centering
\iftoggle{SINGLE_COL}{\begin{subfigure}[b]{.49\linewidth}}{\begin{subfigure}[b]{0.49\linewidth}}
   \includegraphics[width=\linewidth]{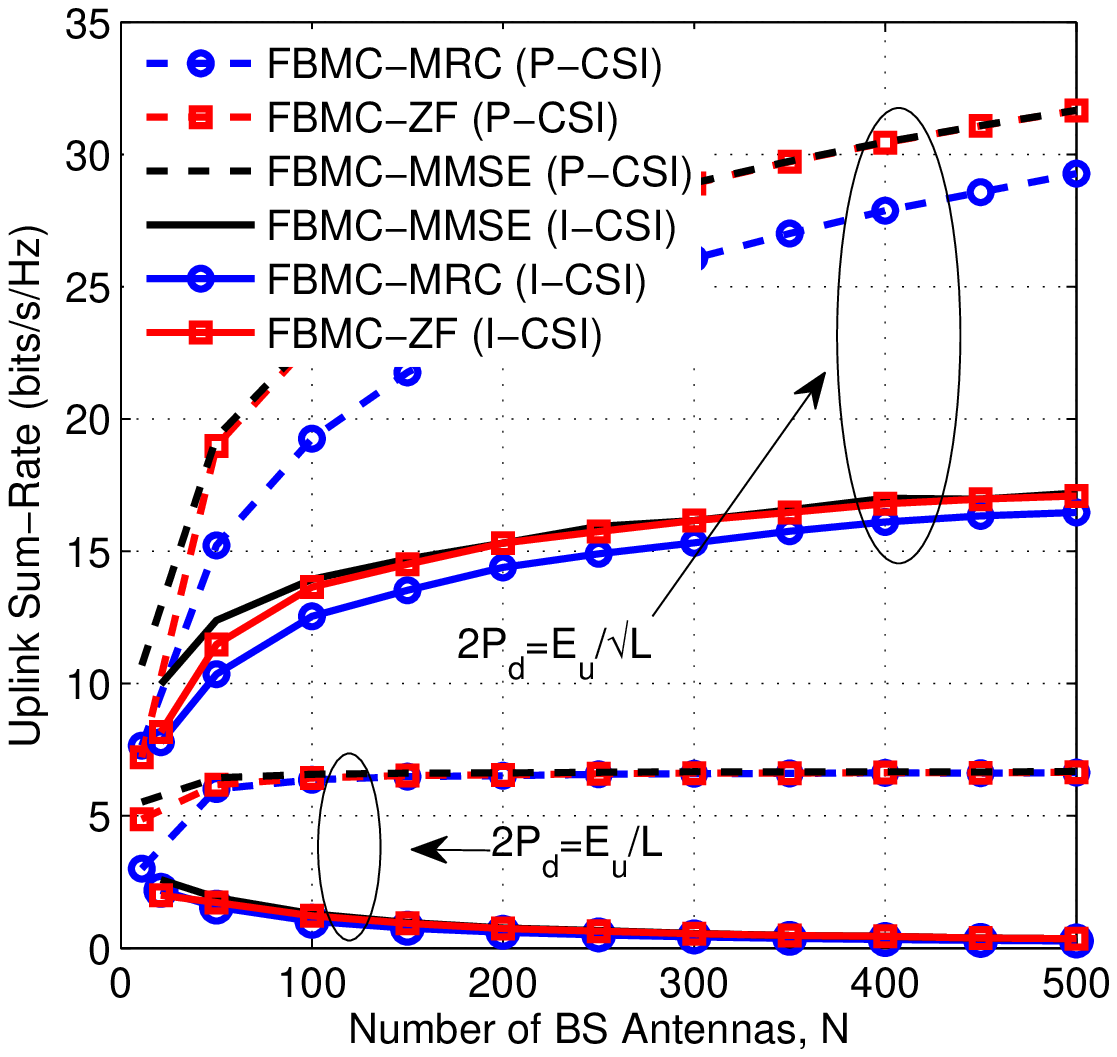}
   \caption{\small }
\label{fig:SC_AR_Vs_nRx_5dB}
\end{subfigure}
\iftoggle{SINGLE_COL}{\begin{subfigure}[b]{.49\linewidth}}{\begin{subfigure}[b]{0.49\linewidth}}
   \includegraphics[width=\linewidth]{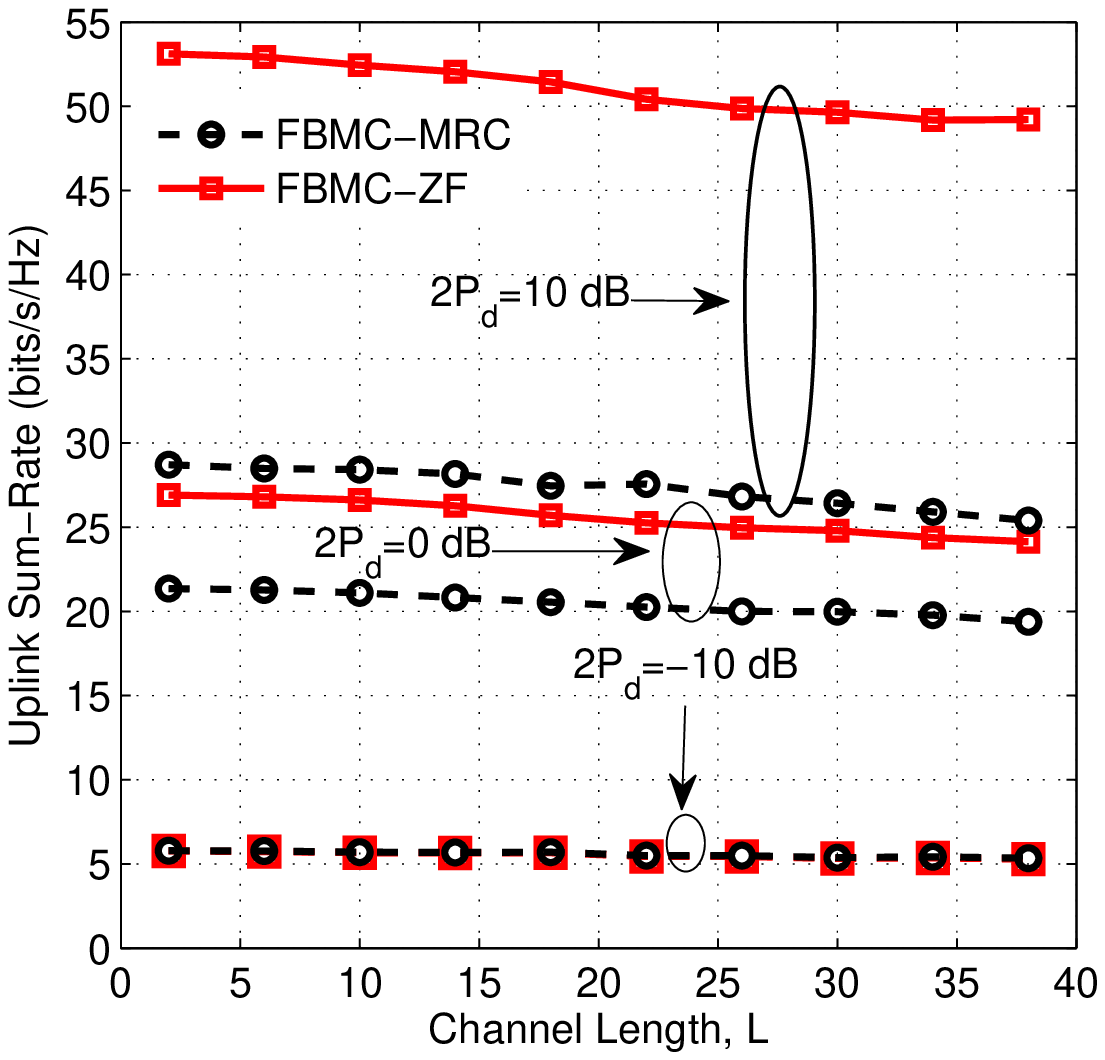}
   \caption{\small }
\label{fig:SC_AR_Vs_Pu_Lh40}
\end{subfigure}
\caption{Uplink sum-rate versus a) number of BS antennas with perfect and imperfect CSI, the reference transmit power per user $E^{u}=5$ dB; and b) the channel length in the presence of imperfect CSI with $N=128$ and $M=64$.}
\end{figure}

Fig.~\ref{fig:SC_AR_Vs_nRx_5dB} verifies the power scaling laws, which are derived in Sections \ref{SC_MU_FBMC_IMCSI} and \ref{SC_MU_FBMC_PCSI} for the single-cell MU massive MIMO-FBMC system employing  the MMSE, ZF and MRC combining at the BS with/ without perfect CSI. The reference power $E^{u}$ per user is fixed at $5$ dB. It is observed that when each of the users scales down its power as $2P_{d}=E^{u}/{N}$ in the presence of perfect CSI at the BS, the uplink sum-rate of all the receivers can be seen to approach a non-zero value. However, for the case of imperfect CSI associated with $2P_{d}=E^{u}/{N}$, the uplink sum-rate of all the receivers approaches zero, as the number of BS antennas increases. On the other hand, with $2P_{d}=E^{u}/\sqrt{N}$, the sum-rate increases without bound with the number of BS antennas for the perfect CSI case. However, for imperfect CSI, the sum-rate converges to a non-zero value.  This study confirms that power scaling laws, similar to OFDM \cite{NgoLM13}, also hold for MU massive MIMO-FBMC systems. Typically, MRC performs better than ZF at low SNR and vice-versa at high SNR, whereas MMSE performs best across the entire SNR range. The same can be observed from Fig.~\ref{fig:SC_AR_Vs_nRx_5dB}, wherein the MRC performs close to the ZF and MMSE receivers for large $N$, because in both the cases the effect of MUI is progressively hidden by the noise since the power is proportional to $1/N$ or $1/\sqrt{N}$.

 The orthogonality in FBMC systems progressively degrades as the channel dispersion increases \cite{RottenbergMHL17}. This happens because the approximation error in the assumption $p[l-i-nM/2]\approx p[l-nM/2]$ for $i\in [0\ L]$ (see paragraph below \eqref{eq:Rec_Sig}) increases with the channel impulse response (CIR) length $L$. Consequently, the detected OQAM symbols are affected by a residual interference \cite[eq. (10)]{lin2009analysis}.  Fig.~\ref{fig:SC_AR_Vs_Pu_Lh40} corroborates this effect, wherein the uplink sum-rate of the FBMC-based MU massive MIMO systems with the MRC and ZF receiver processing is plotted as a function of $L$. It is observed that the uplink sum-rate of both the receivers degrades with an increase of the CIR length $L$ when the transmit power per user is $10$ dB. However, when the transmit power per user is $0$ dB or $-10$ dB, the performance of both the receivers remains unaffected to a large extent. This happens because the residual interference imposed by the increased CIR length is negligible in comparison to the noise  power in the low power regime, which otherwise dominates in the low noise power regime (when the power transmits per user is high).
\begin{figure*}[t!]
\centering
\iftoggle{SINGLE_COL}{\begin{subfigure}[b]{.49\linewidth}}{\begin{subfigure}[b]{0.32\linewidth}}
   \includegraphics[width=\linewidth]{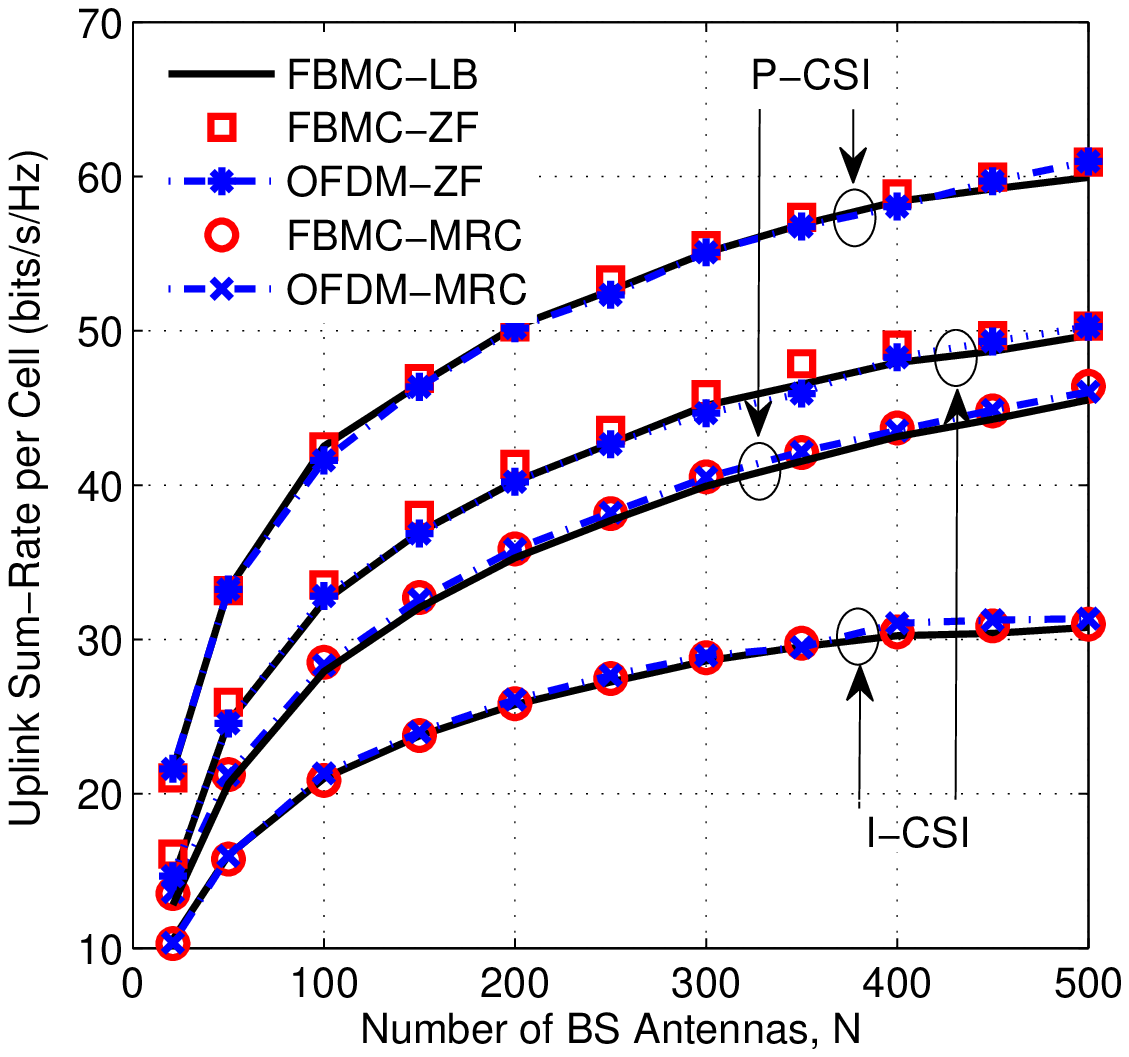}
   \caption{\small }
\label{fig:FBMC_IMCSI_AR_Vs_nRx_10dB}
\end{subfigure}
\iftoggle{SINGLE_COL}{\begin{subfigure}[b]{.49\linewidth}}{\begin{subfigure}[b]{0.32\linewidth}}
   \includegraphics[width=\linewidth]{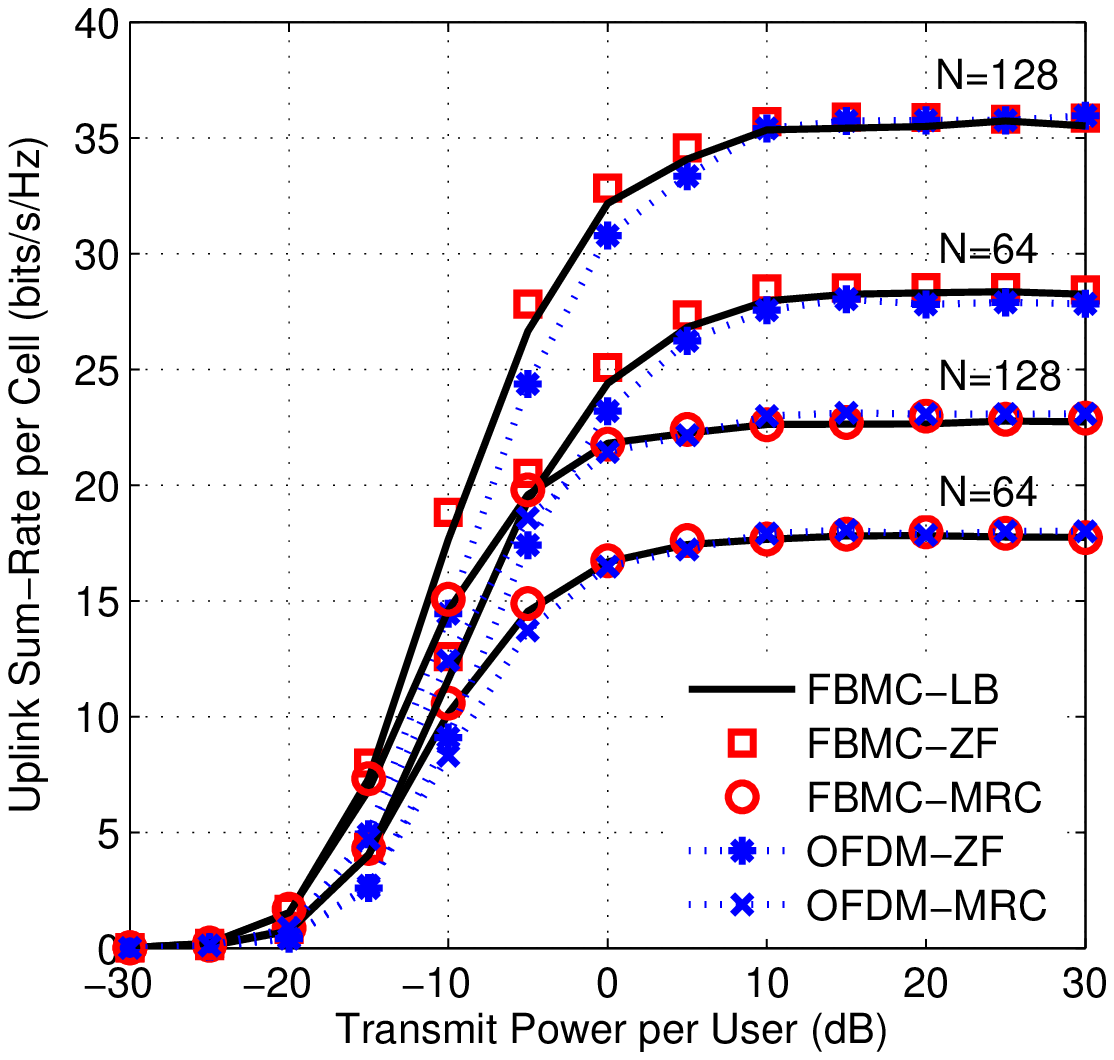}
   \caption{\small }
\label{fig:MC_AR_Vs_Pu}
\end{subfigure}
\caption{Uplink sum-rate per cell versus a) number of BS antennas for perfect and imperfect CSI with  the transmit power per user $2P_{d}=10$ dB; and b) transmit power per user for different number of BS antennas in the presence of imperfect CSI.}
\end{figure*}

\subsection{Multi-Cell Uplink Scenario}
A system having $N_{c}=7$ cells is considered with the radius of each cell set as $r=1000$ meters. It is assumed that $U=8$ single-antenna users are located uniformly at random  in each cell with a radius ranging from $r_{h}=100$ to $1000$ meters. The large-scale fading coefficients obey $\beta^{u}_{n,n}=1$, and for $i\neq n$, they are modelled as $\beta^{u}_{n,i}=z_{i}^{u}/(r_{i}^{u}/r_{h})^{\nu}$. Here $z^{u}_{i}$ is a log-normal random variable for the $u$th user in the $i$th cell with a standard deviation $\sigma_{z}$, $r_{i}^{u}$ is the distance between the $u$th user in the $i$th cell and the BS and $\nu$ is the path loss exponent. The parameters $\sigma_{z}$ and $\nu$ are assumed to be $8$ dB and $3.8$, respectively. The achievable uplink sum-rates per cell at the $\bar{m}$th subcarrier for perfect and imperfect CSI at the BS are defined as $\sum_{u=1}^{U}\mathcal{R}^{u,\text{A}}_{\bar{m},n,\text{P}}$ and $\frac{T_{0}-K}{T_{0}}\sum_{u=1}^{U}\mathcal{R}^{u,\text{A}}_{\bar{m},n,\text{IP}}$, respectively with the corresponding expressions for the lower bounds  as $\sum_{u=1}^{U}\mathcal{\tilde{R}}^{u,\text{A}}_{\bar{m},n,\text{P}}$ and $\frac{T_{0}-K}{T_{0}}\sum_{u=1}^{U}\mathcal{\tilde{R}}^{u,\text{A}}_{\bar{m},n,\text{IP}}$, respectively, where $T_{0}=196$, $\text{A}\in (\text{mrc, zf})$ and the quantities $\mathcal{R}^{u,\text{A}}_{\bar{m},n,\text{P}}$, $\mathcal{R}^{u,\text{A}}_{\bar{m},n,\text{IP}}$, $\mathcal{\tilde{R}}^{u,\text{A}}_{\bar{m},n,\text{P}}$ and $\mathcal{\tilde{R}}^{u,\text{A}}_{\bar{m},n,\text{IP}}$ are described in Section-\ref{MC_MU_MMIMO}.

Fig.~\ref{fig:FBMC_IMCSI_AR_Vs_nRx_10dB} shows the uplink sum-rate per cell versus the number of BS antennas for perfect and imperfect CSI scenarios.  For both the MRC and ZF receivers with perfect and imperfect CSI, the lower-bounds derived in Sections \ref{MC_PCSI} and \ref{MC_IMCSI}  can be seen to closely match the plots obtained via simulation, thus validating the analytical results. It can also be observed that the FBMC-based multi-cell MU massive MIMO system using both the MRC and ZF receivers performs similar its OFDM counterparts.

Fig.~\ref{fig:MC_AR_Vs_Pu} shows the uplink sum-rate per cell versus transmit power per user in the presence of imperfect CSI. The lower-bounds proposed in Sections \ref{MC_PCSI} and \ref{MC_IMCSI} can again be seen in conformance with their respective simulated plots. Since the ZF receiver cancels the interference from the users within the desired cell, it can be seen to outperform the MRC receiver in the high power regime, where the effect of noise is negligible. In contrast to the single-cell case, the interference from the users in other cells saturates the  performance of the ZF receiver. In the low power regime where the noise dominates, the MRC receiver maximizing the SNR performs similar to that of the ZF receiver.
\begin{figure}[t!]
\centering
\iftoggle{SINGLE_COL}{\begin{subfigure}[b]{.49\linewidth}}{\begin{subfigure}[b]{0.49\linewidth}}
   \includegraphics[width=\linewidth]{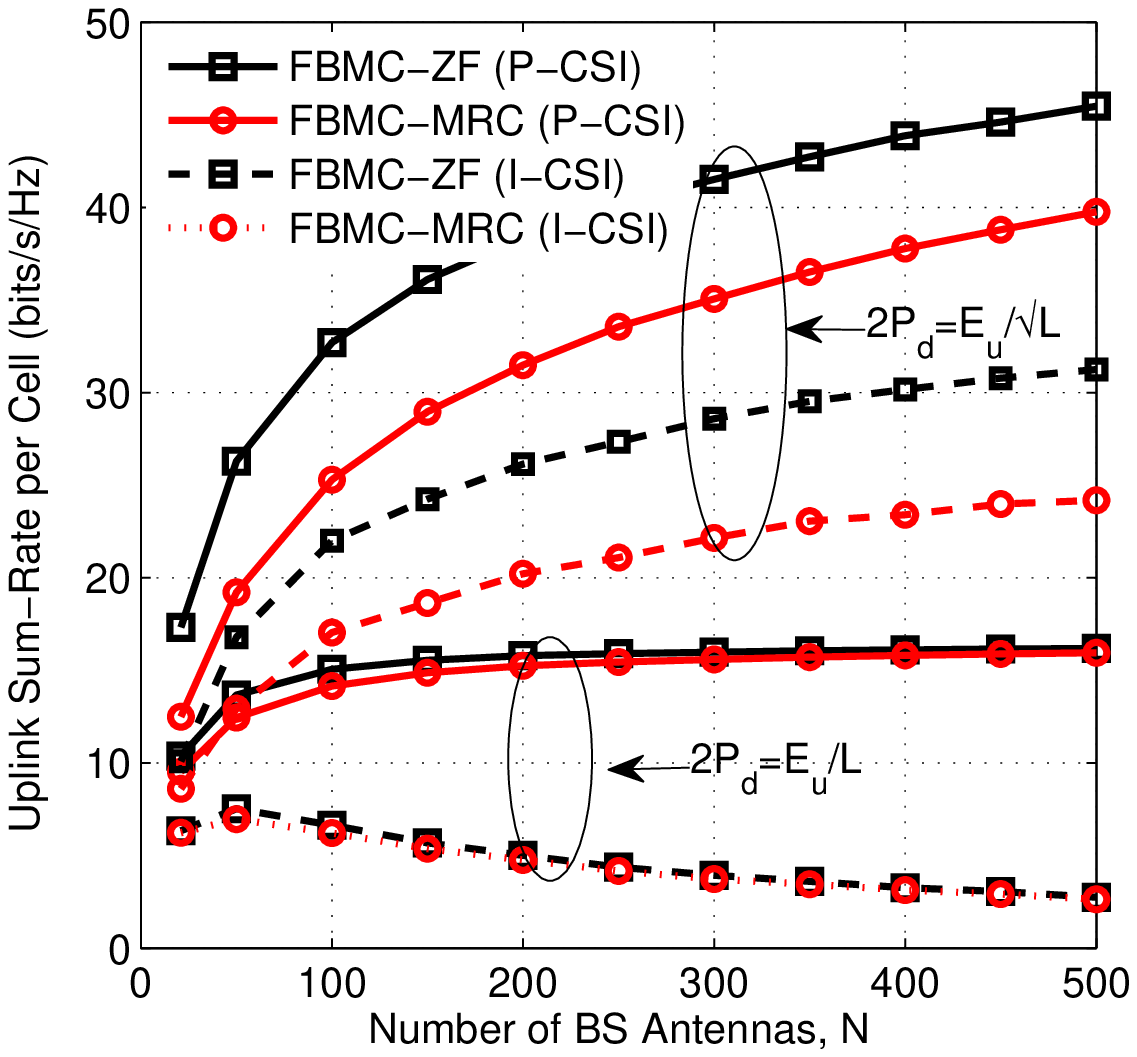}
   \caption{\small }
\label{fig:MC_AR_Vs_nRx_5dB}
\end{subfigure}
\iftoggle{SINGLE_COL}{\begin{subfigure}[b]{.49\linewidth}}{\begin{subfigure}[b]{0.49\linewidth}}
   \includegraphics[width=\linewidth]{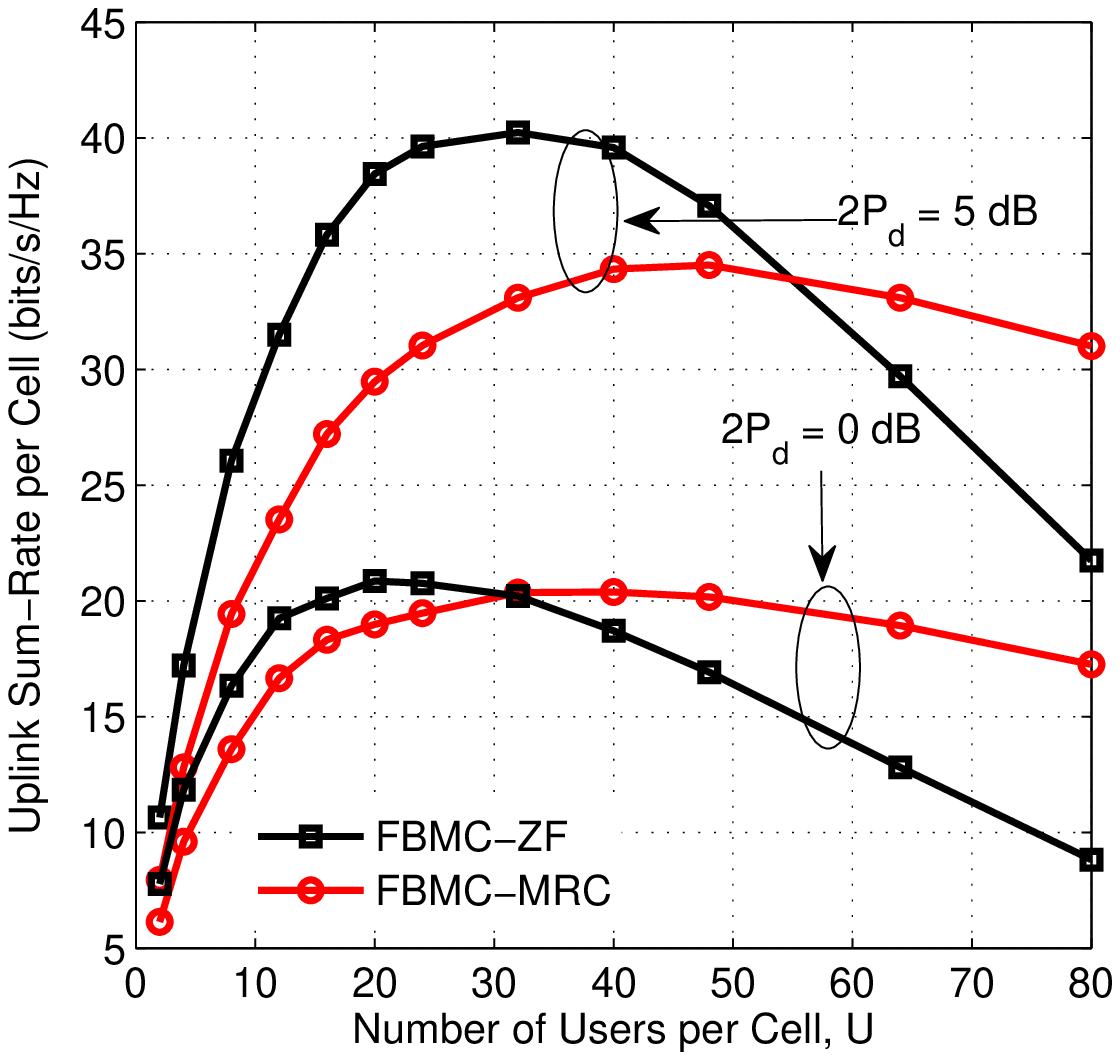}
   \caption{\small }
\label{fig:MC_OFDM_IM_CSI_AR_Vs_nUser}
\end{subfigure}
\caption{Uplink sum-rate versus a) number of BS antennas with perfect and imperfect CSI, the reference transmit power $E^{u}=5$ dB; and b) the number of users per cell $U$ in the presence of imperfect CSI with $N= 128$ BS antennas.}
\end{figure}

Fig.~\ref{fig:MC_AR_Vs_nRx_5dB} confirms the power scaling laws obtained in Section-\ref{MC_MU_MMIMO} for the multi-cell MU massive MIMO-FBMC systems in the presence of perfect and imperfect CSI. It is observed that when the power of each user is proportional to $1/\sqrt{N}$, the uplink sum-rates per cell in the presence of perfect CSI grow without bound, whereas in the presence of imperfect CSI, they approach a non-zero value.  On the other hand, when the transmit power per user is proportional to $1/N$, the uplink sum-rates per cell with perfect CSI converge to a non-zero value, whereas they approach zero in the case of imperfect CSI.


Fig.~\ref{fig:MC_OFDM_IM_CSI_AR_Vs_nUser} portrays uplink sum-rate per cell versus the number of users per cell with imperfect CSI for $128$ BS antennas. In this study, a total power of $5$ dB per cell is divided equally among the users within that cell. This experiment is subsequently repeated  for a total power value of $0$ dB. As the number of users per cell increases, the power per user decreases. As a result, the intra-cell as well as inter-cell interference decreases and the noise effect starts to dominate. Consequently, the performance of the ZF receiver degrades as the number of users increases, and eventually the MRC receiver starts to perform better than the ZF receiver.
\begin{figure}[t!]
\centering
\iftoggle{SINGLE_COL}{\begin{subfigure}[b]{.49\linewidth}}{\begin{subfigure}[b]{0.49\linewidth}}
   \includegraphics[width=\linewidth]{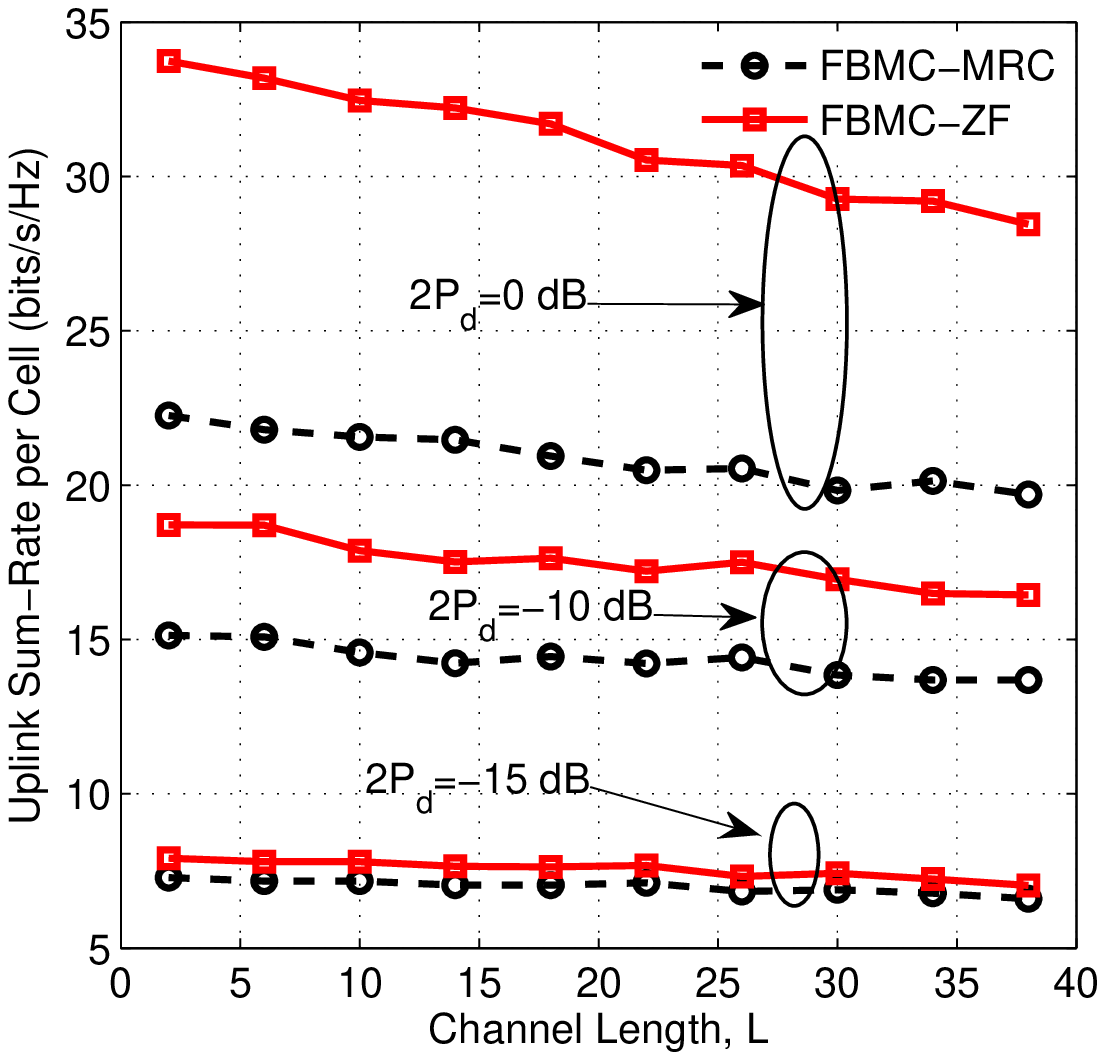}
   \caption{\small }
\label{fig:MC_AR_Vs_Pu_Lh40}
\end{subfigure}
\iftoggle{SINGLE_COL}{\begin{subfigure}[b]{.49\linewidth}}{\begin{subfigure}[b]{0.49\linewidth}}
   \includegraphics[width=\linewidth]{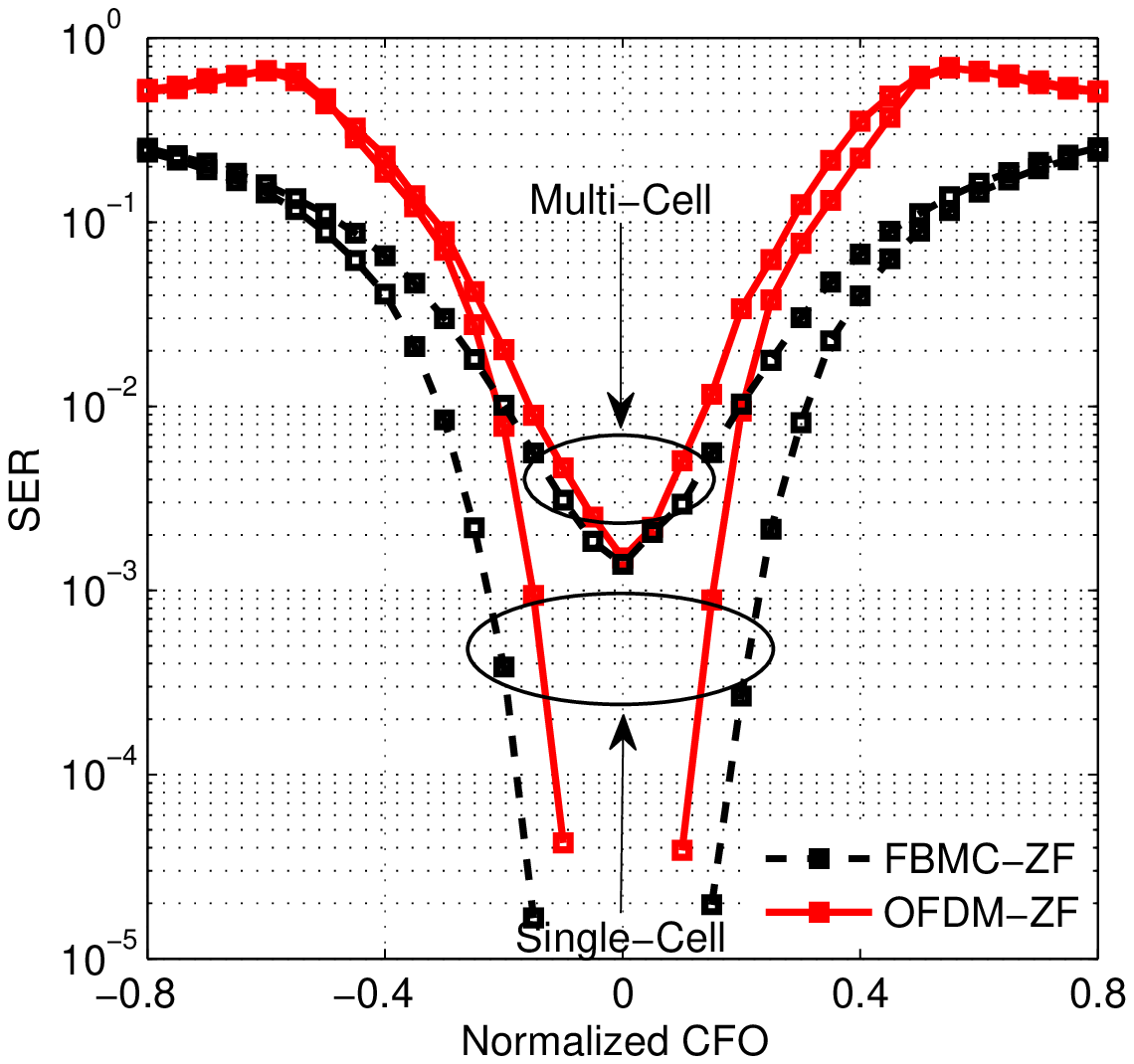}
   \caption{\small }
\label{fig:SC_FBMC_IM_CSI_AR_Vs_nUser}
\end{subfigure}
\caption{a) Uplink sum-rate versus the channel length $L$ in the presence of imperfect CSI with $N=128$ BS antennas and $M=64$ subcarriers. b) SER versus normalized CFO performance of the OFDM and FBMC-based single- and multi-cell massive MIMO systems for BPSK symbols with ZF receiver processing at the base station in the presence of CFO and perfect CSI. Power per user $2P_{d}=-5$ dB, $N=64$, $U=8$, $L=2$. For single-cell: $\beta^{u}=1$ for $1\leq u\leq U$. For multi-cell:  the large scale fading coefficients $\beta^{u}_{n,n}=1\ \forall\ 1\leq u\leq U$, and  for $i\neq n$ $\beta^{u}_{n,i}=0.1\ \forall\ 1\leq u\leq U$.}
\end{figure}

Fig.~\ref{fig:MC_AR_Vs_Pu_Lh40} displays the performance of the ZF and MRC receivers as a function of the channel's delay spread and shows a trend similar to that of Fig.~\ref{fig:SC_AR_Vs_Pu_Lh40}. It is observed that the sum-rate of both the receivers degrades when the transmit power per user is $10$ dB, since the residual interference generated due to the increase in the CIR length dominates in the high-power regime. However, since the system is noise limited in the low-power regime, the channel's delay spread does not significantly impact the performance of both the receives, when the transmit power per user is $-10$ dB or $-15$ dB.

Fig.~\ref{fig:SC_FBMC_IM_CSI_AR_Vs_nUser} shows the SER performance for the OFDM and FBMC-based single- and multi-cell massive MIMO systems in the presence of CFO and perfect CSI at the base station. For simplicity, all the users, in both the cases, are assumed to experience the same amount of CFO. The CFO is normalized to the subcarrier spacing $1/T$. Since the increase of the CFO progressively degrades the subcarrier orthogonality, the poor sinc-shaped frequency localization of the time domain rectangular pulse in OFDM systems results in a significantly higher ICI. On the other hand, FBMC systems due to the well-localized pulse shape (both in frequency as well as in time) experience significantly lower ICI, which makes these systems robust against the CFO. Therefore, the FBMC-based single- and multi-cell massive MIMO systems significantly outperform their OFDM-based counterparts.

\section{Conclusions} \label{Conclusion}
This paper analysed the performance of FBMC signaling in single- and multi-cell MU massive MIMO systems. The lower-bounds and asymptotic expressions derived for the uplink sum-rate with the MRC, ZF and MMSE combining in single- and multi-cell systems with/ without prefect CSI at the BS were seen to coincide with the corresponding simulated sum-rates. It was also demonstrated  that the MU massive MIMO-FBMC systems have a performance similar to that of the CP-OFDM systems.  Furthermore, the power scaling laws, similar to CP-OFDM systems, were seen to hold for the FBMC signaling in the uplink. Future research may present a similar analysis for characterising the performance of FBMC-based massive MIMO systems in time-selective channels. Future research may also analyse the sum-rate of the downlink of FBMC-based massive MIMO systems considering the effect of multi-user precoding, which poses additional challenges.
\appendices
\section{Variance of Intrinsic Interference}\label{Intrf_Cal}
Since the OQAM symbols are i.i.d. zero mean  with power $P_{d}$, from \eqref{eq:Intrf}, one obtains   
\begin{eqnarray}\label{eq:VSP}
\mathbb{\mathbb{E}}[|I^{u}_{\bar{m},\bar{k}}|^{2}]=P_{d}\sum_{(m,k)\in\Omega_{\bar{m},\bar{k}}}^{} \big|\langle\xi\rangle^{\bar{m},\bar{k}}_{m,k}\big|^{2}.
\end{eqnarray}
Using \eqref{eq:basis}, the quantity $\sum_{m=0}^{M-1}\sum_{k\in\mathbb{Z}}^{}\big|\xi^{\bar{m},\bar{k}}_{\bar{m}+m,\bar{k}+k}\big|^{2}$ can be evaluated as
\begin{align}
\iftoggle{SINGLE_COL}{\nonumber}{\nonumber &}\sum_{m=0}^{M-1}\sum_{k\in\mathbb{Z}}^{}\big|\xi^{\bar{m},\bar{k}}_{\bar{m}+m,\bar{k}+k}\big|^{2}=\sum_{n=-\infty}^{+\infty}\sum_{l=-\infty}^{+\infty}p[n]p[l]\sum_{k\in\mathbb{Z}}^{}\iftoggle{SINGLE_COL}{}{\\ &}p[n-kM/2]p[l-kM/2]\sum_{m=0}^{M-1}\exp\{j2\pi m(n-l)/M\}.
\end{align}
For $n-l\neq \alpha_{0}M$ with $\alpha_{0}\in\mathbb{Z}$, the quantity $\sum_{m=0}^{M-1}\exp\{j2\pi m(n-l)/M\}=0$, and it is equal to $M$ when $n-l= \alpha_{0}M$.  Upon employing the above results, one obtains
\begin{align}
\nonumber &\sum_{m=0}^{M-1}\sum_{k\in\mathbb{Z}}^{}\big|\xi^{\bar{m},\bar{k}}_{\bar{m}+m,\bar{k}+k}\big|^{2}=M\sum_{l=-\infty}^{+\infty}\sum_{\alpha_{0}\in \mathbb{Z}}^{}p[l]p[l-\alpha_{0}M]\iftoggle{SINGLE_COL}{}{\\ &\times}\sum_{k\in\mathbb{Z}}^{}p[l-kM/2]p[l-(k+2\alpha_{0})M/2].
\end{align}
Since the prototype pulse $p[l]$ is symmetrical,  it follows that for all $l$ the summation $\sum_{k\in\mathbb{Z}}^{}p[l-nM/2]p[l-(k+2\alpha_{0})M/2]=0$ when $\alpha_{0}\neq 0$, and for $\alpha_{0}=0$, we get $\sum_{k\in\mathbb{Z}}^{}p^{2}[l-nM/2]=2/M$ for all $l$ \cite[Eq. (81)]{siohan2002analysis}. Hence the expression $\sum_{m=0}^{M-1}\sum_{k\in\mathbb{Z}}^{}\big|\xi^{\bar{m},\bar{k}}_{\bar{m}+m,\bar{k}+k}\big|^{2}$ simplifies as
\begin{align}\label{eq:VSP3}
\sum_{m=0}^{M-1}\sum_{k\in\mathbb{Z}}^{}\big|\xi^{\bar{m},\bar{k}}_{\bar{m}+m,\bar{k}+k}\big|^{2}\stackrel{(a)}{=}\dfrac{2M}{M} \sum_{l=-\infty}^{+\infty}p^{2}[l]=2,
\end{align}
where (a) above follows from the fact that the pulse $p[l]$ has unit energy, i.e., $\sum_{l=-\infty}^{+\infty}p^{2}[l]=1$. Since FBMC systems comprise well localized  FT pulse shaping filters, we have
\begin{align}
\nonumber \sum_{(m,k)\in\Omega_{\bar{m},\bar{k}}}^{} \big|\langle\xi\rangle^{\bar{m},\bar{k}}_{m,k}\big|^{2}\approx\sum_{m=0}^{M-1}\sum_{k\in\mathbb{Z}}^{}\big|\xi^{\bar{m},\bar{k}}_{\bar{m}+m,\bar{k}+k}\big|^{2}-\big|\xi^{\bar{m},\bar{k}}_{\bar{m},\bar{k}}\big|^{2}=1.
\end{align}
Upon substituting the above result in \eqref{eq:VSP}, we get the desired result in \eqref{eq:Var_Intr}.
\section{Intrinisc Interference Analysis for Channel Estimation}\label{Inter_Cal_chan}
In order to generalize the analysis, let $z$ zeros are inserted between the adjacent training symbols in Fig.~\ref{fig:frame} to suppress the ISI. Thus, the training symbols are located at the symbol indices $k=i(1+z)$ for $0\leq i\leq K-1$. From (\ref{eq:Intrf}), the intrinsic interference for the $u$th user at these indices can be calculated~as
\begin{align}\label{eq:Intrf_A1}
\iftoggle{SINGLE_COL}{\nonumber}{\nonumber &}I^{u}_{\bar{m},i(1+z)}\iftoggle{SINGLE_COL}{&}{}=\sum_{\substack{(m,k)\in \Omega_{\bar{m},i(1+z)}}} d^{u}_{m,k}\Im\Big\{\sum_{l=-\infty}^{+\infty} p\Big[l-k\dfrac{M}{2}\Big]\\ 
&p\Big[l-i(1+z)\dfrac{M}{2}\Big]e^{j{2\pi}(m-\bar{m})l/M}e^{j(\phi_{m,k}-\phi_{\bar{m},i(1+z)})}\Big\}.
\end{align}
The summation over $(m,k)\in \Omega_{\bar{m},i(1+z)}$ in the above expression can be separated into the following three cases. 1) $m\neq\bar{m}$ and $k\neq i(1+z)$; 2) $m=\bar{m}$ and $k\neq i(1+z)$; and 3) $m\neq\bar{m}$ and $k =i(1+z)$. For the first two cases, $d^{u}_{m,k}=0$ in the neighbourhood $\Omega_{\bar{m},i(1+z)}$ of the FT point $(\bar{m},i(1+z))$. Furthermore, as we move away from this neighbourhood, the quantity $p[l-kM/2]p[l-i(1+z)M/2]\approx 0$  due to the well FT localization of the prototype filter $p[l]$. Therefore, $I^{u}_{\bar{m},i(1+z)}\cong 0$ for the first two cases, and only the third case survives where $m\neq\bar{m}$ and $k =i(1+z)$. Since the training symbols locations are $k=i(1+z)$ for $0\leq i\leq K-1$, the neighbourhood $\Omega_{\bar{m},i(1+z)}$  for the third case comprises the non-zero training symbols $d^{u}_{m,k}$. Thus, the intrinsic interference in (\ref{eq:Intrf_A1}) can be computed as
\begin{eqnarray}\label{eq:Intrf_A2}
 I^{u}_{\bar{m},i(1+z)}=\sum_{ m\neq \bar{m}} d^{u}_{m,i(1+z)}\Im\Big\{\sum_{l=-\infty}^{+\infty} p^{2}\Big[l-i(1+z)\dfrac{M}{2}\Big]e^{j{2\pi}(m-\bar{m})l/M}e^{j(\phi_{m,i(1+z)}-\phi_{\bar{m},i(1+z)})}\Big\}.
\end{eqnarray}
Substituting $l-i(1+z)M/2=l$ and $\phi_{m,k}=(\pi/2)(m+k)-\pi mk$ in (\ref{eq:Intrf_A2}) yields $ I^{u}_{\bar{m},i(1+z)}=\sum_{\substack{ m\neq \bar{m}}} d^{u}_{m,i(1+z)}\langle\xi\rangle^{\bar{m},0}_{m,0}$. Typically, $z=1$ is sufficient to suppress the ISI between the adjacent training symbols due to the well localized FT pulse $p[l]$ in FBMC systems \cite{kofidis2013preamble}. Thus, with $z=1$, we get $ I^{u}_{\bar{m},2i}=\sum_{\substack{ m\neq \bar{m}}} d^{u}_{m,2i}\langle\xi\rangle^{\bar{m},0}_{m,0}$.
\section{Construction of Orthogonal $\mathbf{B}_{\bar{m}}$}\label{Othr_Bm}
As shown in the frame structure in Fig.~1, each user transmits $K$ ($K\geq U$) training symbols on each subcarrier for channel estimation. The construction of the orthogonal $\mathbf{B}_{\bar{m}}$ can be explained using an example with $K=U=2$. Let the first user transmits the OQAM training symbol $d_{m}$ at the training symbol indices $0$ and $2$ on the $m$th subcarrier, the second user on the other hand uses the same preamble, but with reversed signs at the symbol instant $2$. Using this precoding at the users end and the relation $I^{t}_{\bar{m},2i}=\sum_{\substack{ m\neq \bar{m}}} d^{t}_{m,2i}\langle\xi\rangle^{\bar{m},0}_{m,0}$ from Appendix-B, it can be readily verified that the virtual symbols obey $b^{1}_{\bar{m},0}=b^{1}_{\bar{m},2}=b^{2}_{\bar{m},0}=-b^{2}_{\bar{m},2}=b_{\bar{m}}$. Thus, the virtual training matrix $\mathbf{B}_{\bar{m}}$ at the receiver can be obtained as
\begin{eqnarray}\label{eq:T_m}
 \nonumber \mathbf{B}_{\bar{m}}=
\begin{bmatrix}
 {b}_{\bar{m}}&{b}_{\bar{m}} \\
{b}_{\bar{m}}&-{b}_{\bar{m}}
\end{bmatrix}={b}_{\bar{m}}\begin{bmatrix}
 1&1 \\
1&-1
\end{bmatrix}=b_{\bar{m}}\mathbf{A}_{2}.
\end{eqnarray}
It can be seen that $\mathbf{A}_{2}$ is an orthogonal matrix and so is the virtual training matrix $\mathbf{B}_{\bar{m}}$.
\section{Variance of Noise plus Interference}\label{Signal_NPI}
Expanding \eqref{eq:MC_R5} using $\hat{\mathbf{g}}^{u}_{\bar{m},n,j}=\beta^{u}_{n,j}\hat{\mathbf{g}}^{u}_{\bar{m},n,n}$ leads to
\begin{align}\label{eq:MC_B1}
\nonumber &v^{u,\text{mrc}}_{\bar{m},\bar{k},n}=\Re\Bigg\{\sum_{j=1,j\neq u}^{U}(\mathbf{\hat{g}}^{u}_{\bar{m},n,n})^{H}\mathbf{\hat{g}}^{j}_{\bar{m},n,n}{b}^{j}_{\bar{m},\bar{k},n}+\sum_{j=1}^{U}(\mathbf{\hat{g}}^{u}_{\bar{m},n,n})^{H}\iftoggle{SINGLE_COL}{\nonumber}{\\ \nonumber &}\mathbf{e}^{j}_{\bar{m},n,n}{b}^{j}_{\bar{m},\bar{k},n}+\sum_{i=1,i\neq n}^{N_{c}}\beta^{u}_{n,i}\norm{\mathbf{\hat{g}}^{u}_{\bar{m},n,n}}^{2}b^{u}_{\bar{m},\bar{k},i}\iftoggle{SINGLE_COL}{\\ &}{}+\sum_{i=1,i\neq n}^{N_{c}}\sum_{j=1,j\neq u}^{U} \iftoggle{SINGLE_COL}{\nonumber}{\\ \nonumber &}(\mathbf{\hat{g}}^{u}_{\bar{m},n,n})^{H}\hat{\mathbf{g}}^{j}_{\bar{m},n,i}{b}^{j}_{\bar{m},\bar{k},i}+\sum_{i=1,i\neq n}^{N_{c}}\sum_{j=1}^{U}(\mathbf{\hat{g}}^{u}_{\bar{m},n,n})^{H}\mathbf{e}^{j}_{\bar{m},n,i}{b}^{j}_{\bar{m},\bar{k},i}\iftoggle{SINGLE_COL}{}{\\ &}+(\mathbf{\hat{g}}^{u}_{\bar{m},n,n})^{H}\boldsymbol{\eta}_{\bar{m},\bar{k},n}\Bigg\}.
\end{align}
Since the virtual symbol $b^{j}_{\bar{m},\bar{k},i}$, noise vector $\boldsymbol{\eta}_{\bar{m},\bar{k},n}$ and  the error vector $\mathbf{e}^{j}_{\bar{m},n,i}$ are zero mean independent, the variance of the noise-plus-interference term $v^{u,\text{mrc}}_{\bar{m},\bar{k},n}$ is equal to the sum of the variances of the individual terms. Employing \eqref{eq:Prel3}, the property $\mathbb{E}\big[\big((\mathbf{\hat{g}}^{u}_{\bar{m},n,n})^{H}\mathbf{\hat{g}}^{j}_{\bar{m},n,n}{b}^{j}_{\bar{m},\bar{k},n}\big)^{2}\big]=0$ and the second-order statistical properties of the intrinsic interference from \eqref{eq:Var_Intr}, the variance of the first term in the above equation is $P_{d}\sum_{j=1,j\neq u}^{U}\big|(\mathbf{\hat{g}}^{u}_{\bar{m},n,n})^{H}\mathbf{\hat{g}}^{j}_{\bar{m},n,n}\big|^{2}$. Exploiting the same set of properties as above, the variances of the third, fourth and sixth terms in the above equation are evaluated as $P_{d}\big|\big|\mathbf{\hat{g}}^{u}_{\bar{m},n,n}\big|\big|^{4}\sum_{i=1,i\neq n}^{N_{c}}(\beta^{u}_{n,i})^{2}$, $P_{d}\sum_{i=1,i\neq n}^{N_{c}}\sum_{j=1,j\neq u}^{U}\big|(\mathbf{\hat{g}}^{u}_{\bar{m},n,n})^{H}\hat{\mathbf{g}}^{j}_{\bar{m},n,i}\big|^{2}$ and $\frac{\sigma^{2}_{\eta}}{2}\big|\big|\mathbf{\hat{g}}^{u}_{\bar{m},n,n}\big|\big|^{2}$, respectively. Employing the same set of properties again along with \eqref{eq:MC_CE5} and \eqref{eq:MC_CE9}, the variances of the second and fifth terms in the above equation can be computed as
\begin{eqnarray}\label{eq:MC_B2}
\nonumber P_{d}\norm{\mathbf{\hat{g}}^{u}_{\bar{m},n,n}}^{2}
\sum_{j=1}^{U}\dfrac{P_{p}(\gamma^{j}-1)+\sigma^{2}_{\eta}}{P_{p}\gamma^{j}+\sigma^{2}_{\eta}}
\end{eqnarray}
and
\begin{eqnarray}\label{eq:MC_B3}
\nonumber P_d\norm{\mathbf{\hat{g}}^{u}_{\bar{m},n,n}}^{2}
\sum_{i=1,i\neq n}^{N_{c}}\sum_{j=1}^{U}\dfrac{\beta^{j}_{n,i}(P_{p}\gamma^{j}-P_{p}\beta^{j}_{n,i}+\sigma^{2}_{\eta})}{P_{p}\gamma^{j}+\sigma^{2}_{\eta}},
\end{eqnarray}
respectively. The addition of all the above computed variances yields the desired result in~\eqref{eq:App_B1}.

\ifCLASSOPTIONcaptionsoff
  \newpage
\fi

\bibliographystyle{ieeetran}
\bibliography{MU_FBMC}
\end{document}